\documentclass[iop,onecolumn]{emulateapj}

\usepackage{graphicx}  % needed for figures
\usepackage{dcolumn}   % needed for some tables
\usepackage{bm}        % for math
\usepackage{amssymb}   % for math
\usepackage{amsmath}
\usepackage{color}
\usepackage{multirow}
\usepackage{epstopdf} 

\begin{document}

\newcommand\va{v_{\textsc a}}
\newcommand\taua{\tau_{\textsc a}}
\newcommand\gammai{\gamma_i}
\newcommand\ai{a_i}
\newcommand{\bl}[1]{{\color{blue}#1}}

\title{``Ideally" unstable current sheets and the triggering of fast magnetic reconnection}

\author{Anna Tenerani}
\affiliation{EPSS, UCLA, Los Angeles, CA}
\email{anna.tenerani@epss.edu.ucla}
\author{Marco Velli}
\affiliation{EPSS, UCLA, Los Angeles, CA}
\author{Fulvia Pucci}
\affiliation{Dipartimento di Fisica, Universit\`a degli studi di Roma Tor Vergata, Rome, Italy}
\author{Simone Landi}
\affiliation{Dipartimento di Fisica e Astronomia, Universit\`a di Firenze, Florence, Italy}
\author{Antonio Franco Rappazzo}
\affiliation{EPSS, UCLA, Los Angeles, CA}

\begin{abstract}
Magnetic reconnection is thought to be the dynamical mechanism underlying many explosive phenomena observed both in space and in the laboratory, though the question of how fast magnetic reconnection is triggered in such high Lundquist ($S$) number plasmas has remained elusive. It has been well established that reconnection can develop over timescales faster than those predicted traditionally once kinetic scales are reached.  It has also been shown that, within the framework of resistive Magnetohydrodynamics (MHD), fast reconnection is achieved for thin enough sheets via the onset of the so-called plasmoid instability. The latter was discovered in studies specifically devoted to the Sweet-Parker current sheet, either as an initial condition or an apparent transient state developing in nonlinear studies. On the other hand, a fast tearing instability can grow on an ideal, i.e., $S$-independent, timescale (dubbed ``ideal" tearing) within current sheets whose aspect ratio scales with the macroscopic Lundquist number as $L/a\sim S^{1/3}$ -- much smaller than the Sweet-Parker one -- suggesting a new way to approach to the initiation of fast reconnection in collapsing current configurations. Here we present an overview of what we have called ``ideal" tearing in resistive MHD, and discuss how the same reasoning can be extended to other plasma models commonly used that include electron inertia and kinetic effects. We then discuss a scenario for the onset of ``ideal'' fast reconnection via collapsing current sheets and describe a quantitative model for the interpretation of the nonlinear evolution of ``ideally" unstable sheets in two dimensions.
\end{abstract}

%\begin{PACS}
%Authors should not enter PACS codes directly on the manuscript, as these must be chosen during the online submission process and will then be added during the typesetting process (see http://www.aip.org/pacs/ for the full list of PACS codes)
%\end{PACS}
\maketitle

\section{Introduction}

Magnetic reconnection is a process whereby magnetic energy is converted locally into particle heat and kinetic energy via some mechanism of effective magnetic dissipation that allows for a change of magnetic field line connectivity. Magnetic reconnection is ubiquitous in space and laboratory plasmas, and is believed to be at the heart of many observed  phenomena, such as solar flares~\citep{masuda,su_2013}, geomagnetic substorms~\citep{angelo}, and  sawtooth crashes in tokamaks~\citep{kad,yamada_1994}. Apart from these transient events, reconnection is also invoked in coronal heating models in different extensions of the  nanoflare scenario~\citep{parker_1988, rappazzo_2008}, and plays a fundamental role during dynamo processes in primordial galaxy clusters~\citep{scheko}. 

Several phenomena in which magnetic reconnection is thought to take place exhibit an explosive character, in the sense that  magnetic energy can be stored over a long period of time, and then suddenly released on a timescale comparable with the macroscopic ideal Alfv\'en time $\taua=L/\va$, where $L$ is the macroscopic length of the system and $\va=B_0/\sqrt{4\pi\rho_0}$ the Alfv\'en speed defined through typical values of magnetic field magnitude $B_0$ and plasma mass density $\rho_0$. For many years, studies of reconnection stumbled on understanding how fast reconnection is triggered. The major difficulty came from the fact that the traditional models of reconnection stemming from the original Sweet-Parker mechanism~\citep{sweet,parker_57} or from the instability of macroscopic current sheets~\citep{FKR}, dating back to the sixties, were clearly inadequate to explain the observed sudden release of magnetic energy, as such models predict magnetic reconnection timescales -- scaling with a positive power of $S$, where $S=L\va/\eta\simeq10^6-10^{14}$ is the macroscopic Lundquist number, $\eta$ being the magnetic diffusivity -- that are far too long to be of any consequence. {Several attempts involved locally enhancing the value of diffusivity by invoking anomalous resistivities to make the Sweet-Parker current layer transition to the fast, steady-state Petschek configuration~\citep{petschek}. However, as discussed, e.g., in~\citet{shibata}, these also require the formation of extremely small scales in the plasma. }

{ The aim of the present review is to discuss how the difficulty of apparently slow reconnection has been overcome, following the works of  \cite{bisk_1986}, studies of the plasmoid instability \citep{lou_2007},  and  the fractal reconnection scenario introduced by  \cite{shibata},}
with emphasis on research carried out by the present authors and in particular on the ``ideal" tearing scenario introduced in~\cite{pucci}. Magnetic reconnection has been the subject of intense research both theoretically and observationally, and in very different physical as well as astrophysical contexts. A complete review on the subject would go well beyond the purpose of the present paper. Here we have tried to include most of the recent papers pertaining to fast reconnection, though the discussion may be brief. A longer review can however be found for instance in~\citet{yamada_2010}.  

Over the past decades a vast body of literature has focussed on what might accelerate {  reconnection speed} up to realistic values. For the most part,  these works approach the problem by studying (two-dimensional) magnetic reconnection at a single X-point, usually imposed by deforming an initially thick current sheet. The ensuing dynamics at the X-point  (sometimes called the developmental phase) eventually leads to an inner current sheet (or diffusion region), in which reconnection is studied by assuming a steady-state (or asymptotic phase) is reached. Two major scenarios for onset of fast reconnection have emerged in this way (see also~\cite{dau_2012, cassak_2013,huang_2013,lou_2015}).
\begin{figure}
\begin{center}
\fbox{{\includegraphics[width=0.6\textwidth]{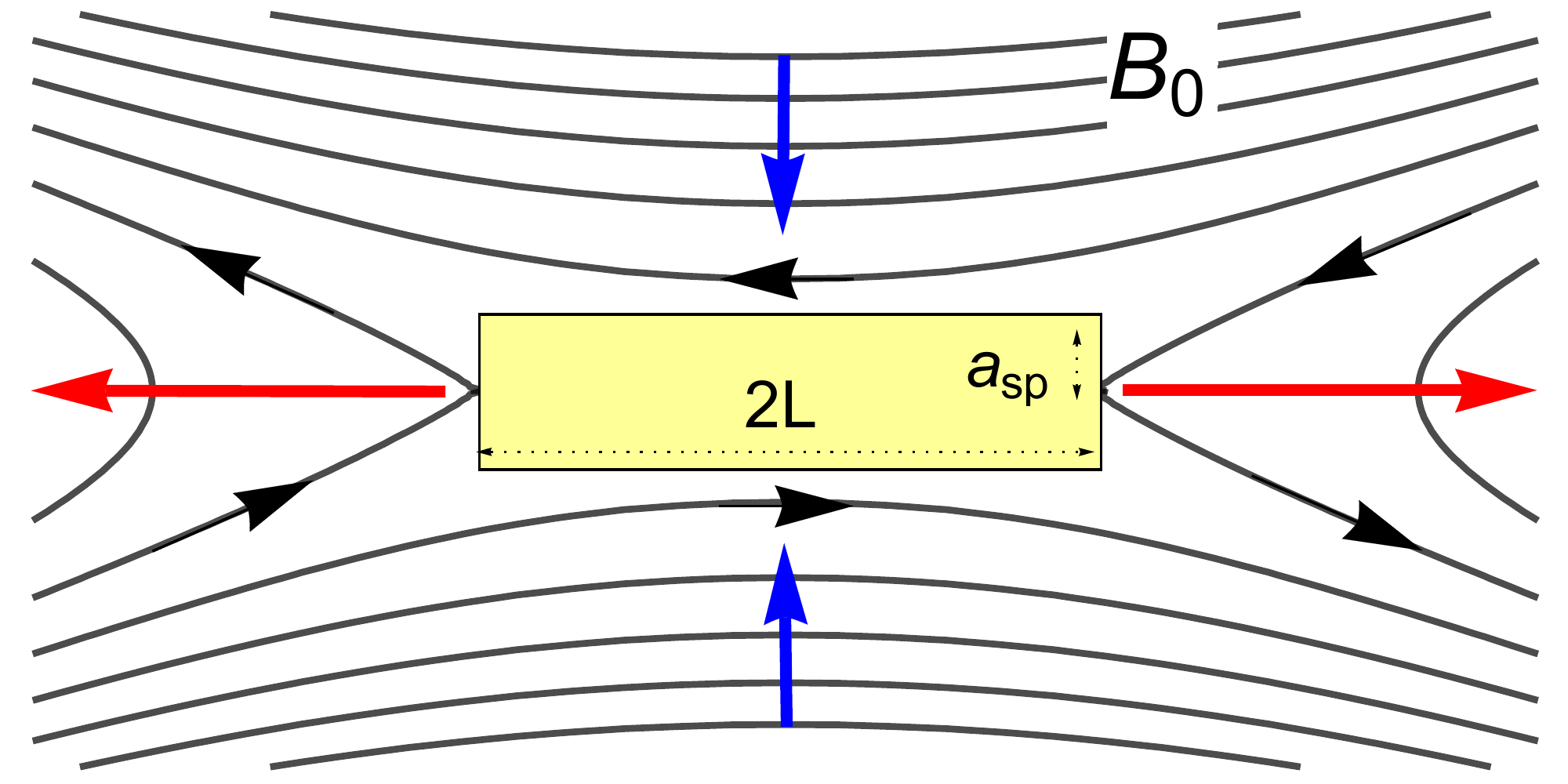}}}
\caption{Sweet-Parker model. The diffusion region, in yellow, has an inverse aspect ratio $a_{\text{\sc sp}}/L$. Colored arrows represent the plasma flow into and outward the diffusion region.}
\label{SP}
\end{center}
\end{figure}

In resistive Magnetohydrodynamics (MHD), numerical simulations show that slowly reconnecting current sheets reminiscent of the Sweet-Parker model (SP) arise from the X-point. The basic SP configuration is shown in Fig.~\ref{SP}: the current sheet has an inverse aspect ratio $a_{\text{\sc sp}}/L\sim S^{-1/2}$, maintained by a continuous  inflow of plasma  at speed $u_{in}\ll\va$ which convects the upstream magnetic field ${\bf B}_0$ into the diffusion region, and by an outflow at speed $u_{out}\simeq \va$ which drags the reconnected magnetic field lines outwards, along the sheet; the resulting  Alfv\'en-normalized reconnection rate is $u_{in}/u_{out}\sim a_{\text{\sc sp}}/L\sim S^{-1/2}$, assuming the steady-state configuration remains stable. Once computational power allowed to study systems at larger values of $S$ ($S\gtrsim 10^4$),  it became clear that SP-like sheets become unstable to tearing ~\citep{bisk_1986, bisk_2000}, the latter inducing the growth of a large number of magnetic islands (plasmoids). In this regard, linear stability analysis shows that SP sheets are highly unstable to a super-Alfv\'enic  tearing instability -- with a growth rate scaling with a positive power of $S$, $\gamma_{\text{\sc  sp}}\taua\sim S^{1/4}$ -- as convincingly proved by~\citet{lou_2007,lou_2013}.  { Though such an instability demonstrates the possibility of fast reconnection already in resistive MHD, \cite{pucci} pointed out that the growth rate, which diverges in the ideal limit $S\rightarrow\infty$, poses consistency problems, since magnetic reconnection is prohibited in ideal MHD.  In other words, they questioned the realizability of current sheets with excessively large aspect ratios, such as SP,  which, when taking the ideal limit starting from resistive MHD, reach an infinite growth rate, i.e., become unstable on timescales which at large enough $S$ are much shorter than any conceivable dynamical time required to set-up the corresponding configuration.  Though this issue of reconnection speed was implicitly recognized in \cite{lou_2007} and many other works,~\citep{cassak_2009, samta,batta_2009, huang_2010,uzdensky_2010,lou_2012,  ni_2015}  the discussion of the onset of fast tearing has mostly remained anchored to the Sweet-Parker sheet framework.}  

Alternatively,  it has been suggested that if ion-scales are of the order of, or larger than, the thickness of the SP sheet, then two-fluid effects enhance the reconnection rate via what is usually called Hall-mediated reconnection. It was already known that the Hall term in Ohm's law   { increases the reconnection speed}  because of the excitation of dispersive waves, e.g., whistler waves~\citep{terasawa,mandt,biskamp_1995,rogers}. To be more specific, it was argued, on the basis of  numerical simulations, that  large reconnection rates should be achieved during the  nonlinear asymptotic phase,  regardless of the  mechanisms allowing reconnection (e.g., resistivity or electron inertia in collisionless reconnection), essentially because the Hall term modifies the structure of the diffusion region that  becomes localized at  scales of the order of the ion inertial length~\citep{shay_1999, shay_2004, cassak_2005, drake_2008, shepherd_2010}.  However, there is still no general agreement on whether and how the reconnection rate depends on the system size $L$ and plasma parameters such as the ion and electron inertial length $d_i$ and $d_e$ or resistivity, in the presence of the Hall term~\citep{porcelli_2002, batta_2005}. More recent numerical results from PIC simulations have also cast  doubt on the necessity of exciting dispersive waves to reach higher reconnection rates~\citep{liu_2014}. The robustness of the steady-state configuration { reached during the asymptotic phase, and  seen in several numerical studies~\citep{birn_2001,shepherd_2010}, has   been called into question by both PIC and Hall-MHD simulations employing open boundaries or larger system sizes~\citep{dau_2006,klimas,huang_2011}. These works provide some evidence that a final steady-state configuration  may not always exist in Hall reconnection, and that the thin sheet constituting the diffusion region tends to stretch along the outflow direction until it becomes  unstable to generation of secondary plasmoids. This would point to a strong analogy with the dynamics of thin current sheets found in resistive~MHD.}

Linear analysis of tearing mode in resistive MHD proves that there exists a critical current sheet with an  inverse aspect ratio $a_i/L\sim S^{-1/3}$~\citep{pucci}, hence much larger than the SP one, that separates slowly reconnecting sheets  (growth rate scaling with a negative power of $S$), from those exhibiting  super-Alfv\'enic plasmoid instabilities (growth rate diverging with $S$), and that this has the proper convergence properties to ideal MHD: critical current sheets are unstable to a tearing mode growing at a rate {\it independent} from $S$ and, in this  sense, the instability is ``ideal". The existence of  ``ideal" tearing therefore implies the impossibility of constructing any configuration corresponding to  sheets thinner than critical, such as the paradigmatic Sweet-Parker sheet, suggesting a different route to the triggering of fast reconnection. At the same time, as we shall see, reasoning in terms of the rescaling arguments provides a clear predictive pathway to the critical events involved in a dynamics that will trigger fast reconnection, { providing an alternative framework within which the many and different results previously obtained in simulations may be reinterpreted.}  It is worth mentioning that one of the major difficulties in this kind of numerical simulations concerns achieving sufficiently large Lundquist numbers. At the intermediate Lundquist numbers usual to MHD simulations, say, $S\simeq10^4-10^6$ the SP sheet is at  most 10 times thinner than the critical current sheet predicted by the ``ideal" tearing theory. Moreover, the possible presence of plasma flows along the current sheet tends to stabilize the tearing mode~\citep{bulanov}, inducing  the formation of more elongated current sheets, that is, of layers having inverse aspect ratios smaller than $S^{-1/3}$ (in the resistive case). Therefore, with the Lundquist number not sufficiently large, the distinction between ``ideal" tearing framework and SP-plasmoids might be hard to observe, though the departure from the SP-plasmoid framework should become increasingly obvious with increasing $S$. 

 Throughout this review we summarize and complement recent results stemming from the ``ideal" tearing idea, obtained from both linear theory and nonlinear numerical simulations, providing a coherent perspective on recent studies and their relation to previous models: ``ideal" tearing can explain the trigger of fast reconnection occurring on critically unstable current sheets and can provide a guide -- at least in two dimensions -- to the nonlinear evolution with a model that describes the different stages of its  evolution. Though we focus mainly on the resistive MHD description of the plasma, the ``ideal" tearing idea can be extended to other regimes, such as two fluid, Hall-MHD and so on to completely kinetic ones, providing a unified and self-consistent framework for the onset of fast reconnection. In this paper we briefly discuss the application to simple kinetic models of  collisionless reconnection. 

The paper is organized as follows: in Section~2 we summarize the  theory of the tearing mode instability in its traditional form, in order to prepare the ground for the ``ideal" tearing, that we discuss in Section~3; in Section~4 we approach the problem of the trigger of fast reconnection and we propose a new scenario relying on  the ``ideal" tearing; in Section~5 we show that the trigger of fast reconnection via ``ideal" tearing can describe the evolution of a disrupting current sheet during its different nonlinear stages, characterized by a recursive sequence of tearing instabilities, and we compare our results with previous existent models in Section 6; a final summary and discussion  are deferred to Section~7. For the sake of clarity, we mention here that in Section 2 and 3 we show results that have been obtained from linear theory by solving numerically the  system of ordinary differential equations of the eigenvalue problem of the tearing mode, with an adaptive finite difference scheme based on Newton iteration~\citep{lentini};  in Section 4 and 5 we show results obtained from 2 and 1/2 dimensional, fully nonlinear resistive MHD simulations. 

\begin{figure}
\begin{center}
\fbox{{\includegraphics[width=0.6\textwidth]{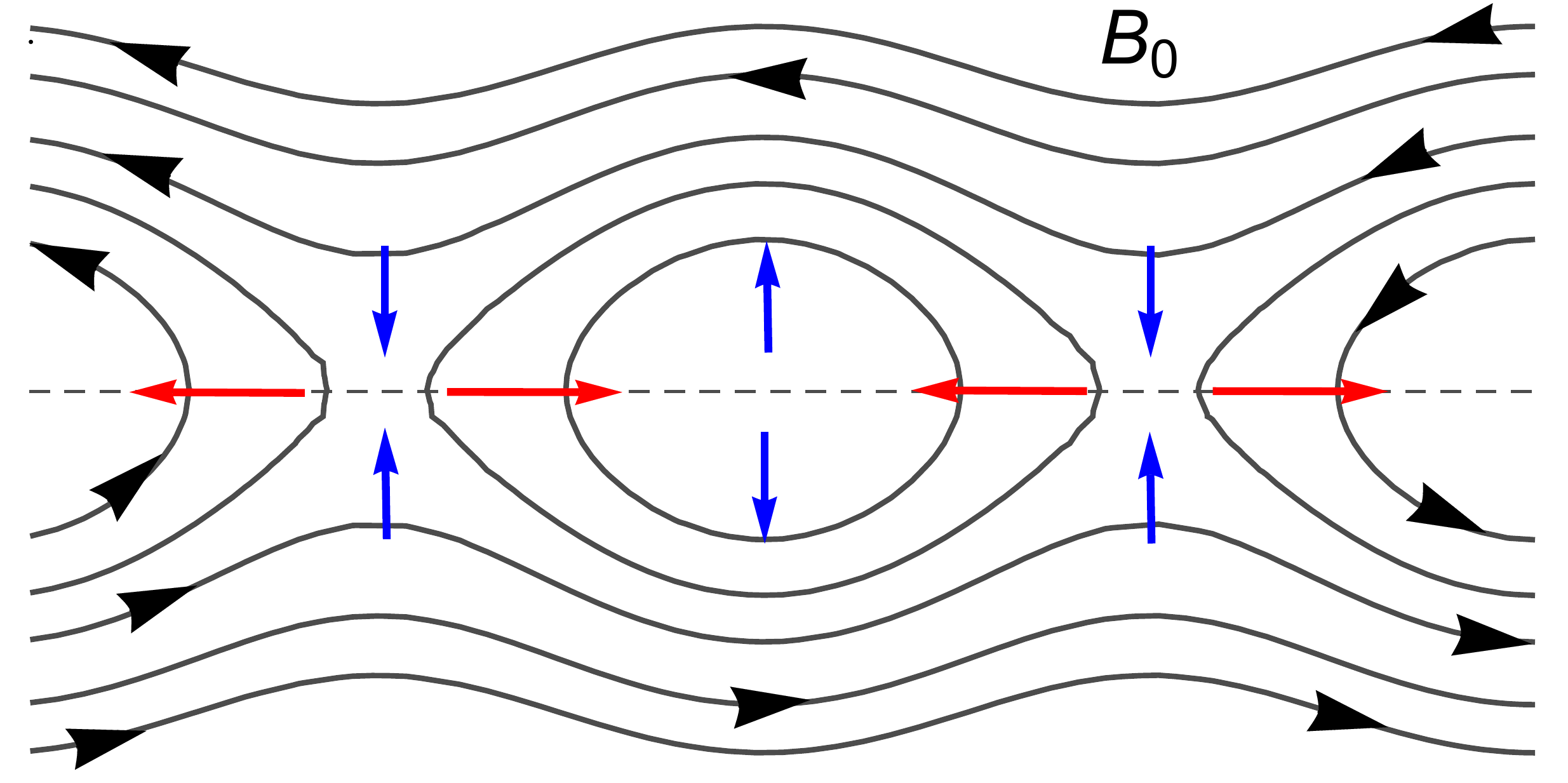}}}
\caption{Tearing mode instability. Colored arrows represent the perturbed plasma flows into and outward the X-point.}
\label{tea}
\end{center}
\end{figure}

\section{Background: the tearing mode instability}
\label{tearing}
Magnetic reconnection can arise spontaneously within current sheets as the outcome of an internal tearing instability  when the non-ideal terms in Ohm's law, resistivity in our case, are taken into account.  Tearing modes  have been extensively studied in the simpler case of an infinite (one-dimensional) sheet both in the linear~\citep{FKR} and nonlinear~\citep{rutherford,wael} regime. Below, we briefly summarize  the basic properties of the instability, obtained from linear analysis in the resistive MHD framework. We assume, as usually done, an incompressible, homogeneous density plasma. 

Tearing modes are long-wavelength modes, that is, unstable perturbations have wavelengths larger than  the shear length (or thickness) $a$ of the equilibrium magnetic field ${\bf B}$. Such unstable modes  lead to the growth of magnetic islands via reconnection at the magnetic neutral line, or, more generally, on ``resonant" surfaces where  ${\bf k\cdot B}=0$,  ${\bf k}$ being the wave vector of the perturbation along the sheet (Fig.~\ref{tea}). Instability grows on a timescale $\sim1/\gamma$ that is intermediate between the dissipative time of the equilibrium magnetic field, $\tau_\eta=a^2/\eta$, and the Alfv\'en crossing time $\bar\tau=a/\va$ based on the thickness $a$. It is worth noting  that since  MHD  does not have any intrinsic scale,  lengths are traditionally normalized to the thickness $a$ of the sheet -- differently from the SP model in which, instead, $L$  is the macroscopic length -- whereas time is normalized to $\bar \tau$. For the sake of clarity, we therefore label with an over-bar the quantities defined through  the current sheet  thickness, $\bar\tau= a/\va$ and $\bar S=a\va/\eta$.

\begin{figure}
\begin{center}
\includegraphics[width=0.47\textwidth]{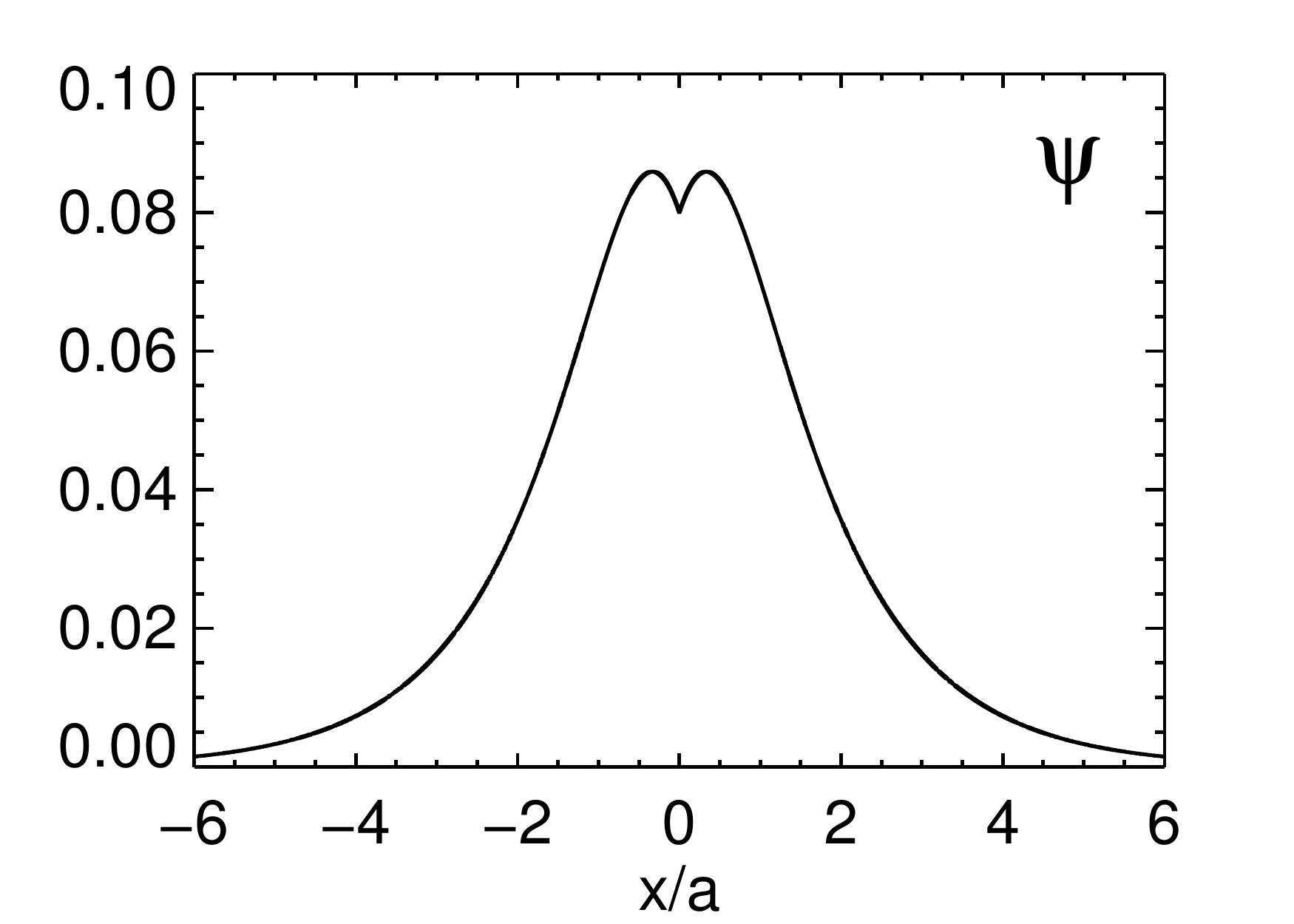}\quad
\includegraphics[width=0.47\textwidth]{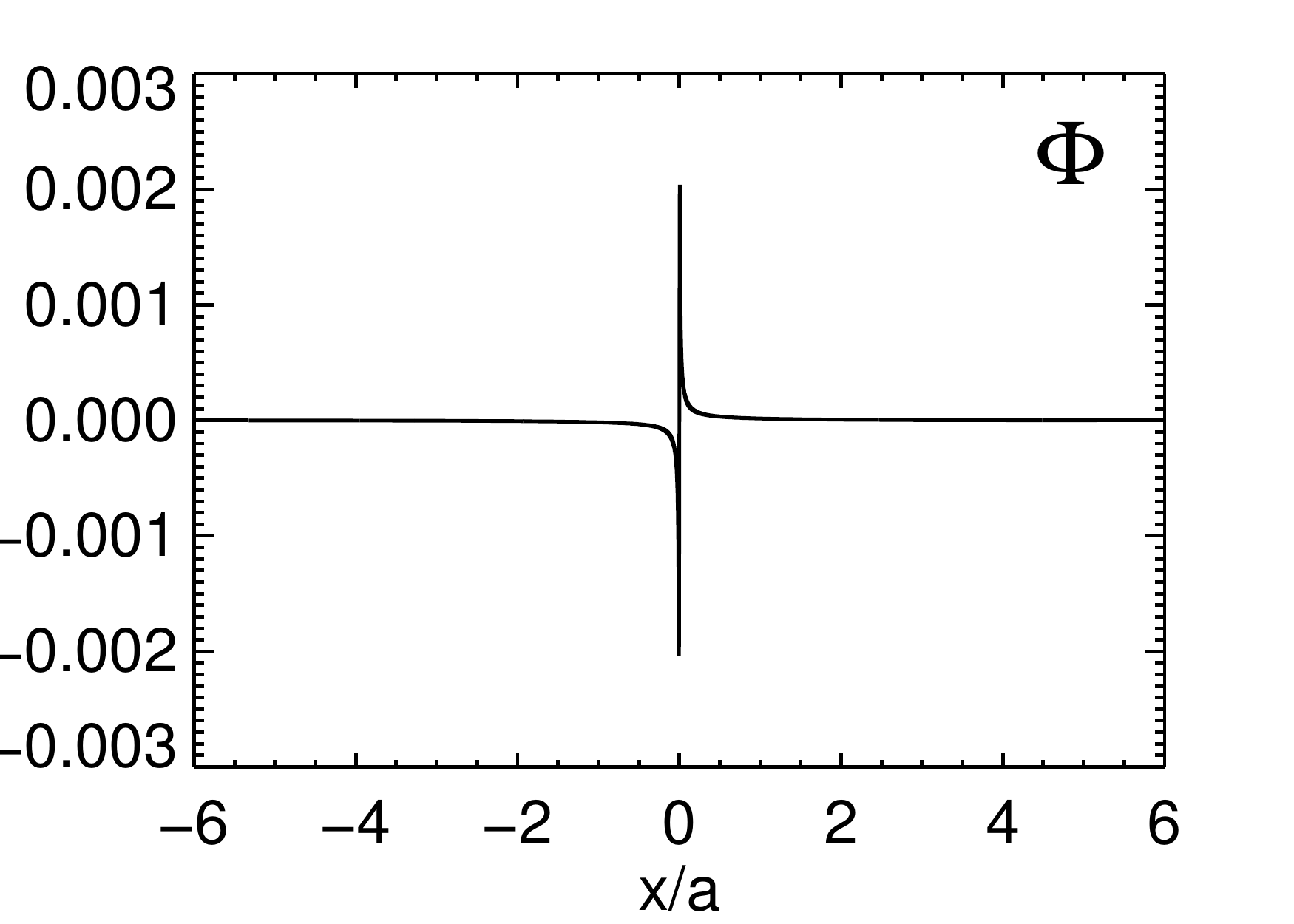}
\caption{Tearing eigenfunctions for $S=10^6$ and $ka=0.8$ (small $\Delta^\prime$): magnetic field flux function $\psi$ (left) and velocity field stream function $\phi$ (right).}
\label{modes}
\end{center}
\end{figure}

Since resistivity is negligible (or rather $\bar S^{-1}\ll1$)  everywhere except close to the resonant surface, tearing modes exhibit  a quasi-singular behavior in a small boundary layer of thickness $\delta$, the inner diffusion region,  in which the perturbed magnetic and velocity fields exhibit sharp gradients, and where reconnection can take place. Out of the  inner region, where resistivity can be neglected,  perturbations smoothly decay to zero far from the resonant surface. An example of the eigenfunctions of the magnetic flux  and velocity stream functions $\psi$ and $\phi$ at $\bar S=10^6$ is shown in Fig.~\ref{modes}. In this case, the plotted eigenfunctions correspond to a Harris current sheet equilibrium, which is the one considered most in the literature and has a magnetic field profile ${\bf B}=B_0\tanh(x/a){\bf\hat y}$. In perfectly antisymmetric equilibria, as the one chosen here, the eigenfunctions have a well-defined parity: the magnetic flux function is symmetric in $x$, being proportional to the reconnected magnetic field component  at the neutral line, whereas the velocity stream function,  proportional to the perturbed plasma velocity perpendicular to the sheet, is antisymmetric. As can be seen in the right panel of Fig.~\ref{modes}, the plasma strongly accelerates while flowing into the inner layer, with the stream function  ideally diverging as $\phi/(a\va)\sim a/x$ for $|x/a|\rightarrow0$. Resistivity  becomes non-negligible close to the neutral line, regularizing the singularity: inside the inner layer, the plasma decelerates  to the stagnation point, where it  deflects outwards along the sheet, while the magnetic field diffuses and reconnects.

Instability is fed by the equilibrium current gradients which enter through the parameter $\Delta^{\prime}(k)$, or for brevity $\Delta^\prime$. The latter is defined as the discontinuity of  the logarithmic derivative of the outer flux function when approaching the singular layer, and is  a measure of the free energy of the system. The $\Delta^\prime$ parameter defines the instability threshold condition, instability occurring only when $\Delta^\prime>0$~\citep{FKR,adler}. For a Harris current sheet $\Delta^\prime a=2[(ka)^{-1}-ka]$, so that the unstable modes {have wavevector satisfying}  $ka<1$. 

The $\Delta^\prime$ parameter also controls the linear evolution. There are indeed two regimes that describe the unstable spectrum depending on the value of $\Delta^\prime\delta$. One  is the so called constant-$\psi$  regime, traditionally referred to as tearing mode or small~$\Delta^\prime$ regime~\citep{FKR}. The other one is the non constant-$\psi$ regime, or large~$\Delta^{\prime}$, also known as resistive internal kink~\citep{coppi_1976}. The difference between these two regimes is in the ordering of the derivatives of $\psi$ within the inner layer that leads to two different limiting cases of the dispersion relation~{\citep{ara_1978}}: in the small $\Delta^\prime$ regime  $\psi^{\prime}\sim\psi\,\Delta^\prime$ and $\psi^{\prime\prime}\sim\psi\,\Delta^\prime\delta^{-1}$, that is, in the inner region $\psi$ is roughly constant and can be approximated by $\psi(0)$, provided $\Delta^\prime\delta\ll1$; at larger values of $\Delta^\prime$ this approximation breaks down as $\psi$ has stronger gradients, $\psi^{\prime}\sim\psi\,\delta^{-1}$ and $\psi^{\prime\prime}\sim\psi\,\delta^{-2}$.  In particular, the growth rate $\gamma$ and the inner layer $\delta$ scale  in these two regimes as
\begin{equation}
\gamma\bar\tau\sim \bar S^{-3/5} (ka)^{2/5}( \Delta^\prime a)^{4/5},\quad\frac{\delta}{a}\sim\bar S^{-2/5}\quad\text{($\Delta^\prime\delta\ll 1$)},
\label{small}
\end{equation}
\begin{equation}
\gamma\bar\tau\sim \bar S^{-1/3}\left(  ka \right)^{2/3},\quad\quad\frac{\delta}{a}\sim\bar S^{-1/3}\quad\text{($\Delta^\prime\delta\gg 1$)}.
\label{large}
\end{equation}
{The expressions above can also be found as special cases of a  general dispersion relation valid for any given value of the parameter $\Delta^\prime$~\citep{peg86}.

Roughly speaking, the small and the large $\Delta^\prime$ regimes have wavevectors lying to the right and to the left of the fastest growing mode $k_m$, respectively. For example, in the case of a Harris sheet the two regimes correspond  to a region in $k$-space $k_ma<ka<1$ in which the growth rate decreases with $k$  (small $\Delta^\prime$) and another one  $0<ka<k_ma$ where the growth rate increases with $k$ (large $\Delta^\prime$). The scaling relations for the fastest growing mode can therefore} be obtained by matching the two regimes~\citep{batta_2009,lou_2013,ideal_de}. Since the expressions for $\gamma\bar\tau$ given in eqs.~(\ref{small})--(\ref{large}) should coincide at the fastest growing mode,  the wavevector $k_m$ can be obtained by equating the right-hand-side of the growth rate in the small and large $\Delta^\prime$ regimes, respectively. The scaling of the maximum growth rate $\gamma_m$ follows directly  by either the small or the large $\Delta^\prime$ growth rate at $k=k_m$, leading to
\begin{equation}
 k_ma \sim \bar S^{-1/4},\quad\quad  \gamma_m\bar\tau\sim \bar S^{-1/2},\quad\quad\frac{\delta_m}{a}\sim \bar S^{-1/4}\quad\text{(Fastest growing mode)}.
\label{fgm}
\end{equation}

\begin{figure}
\begin{center}
\includegraphics[width=0.7\textwidth]{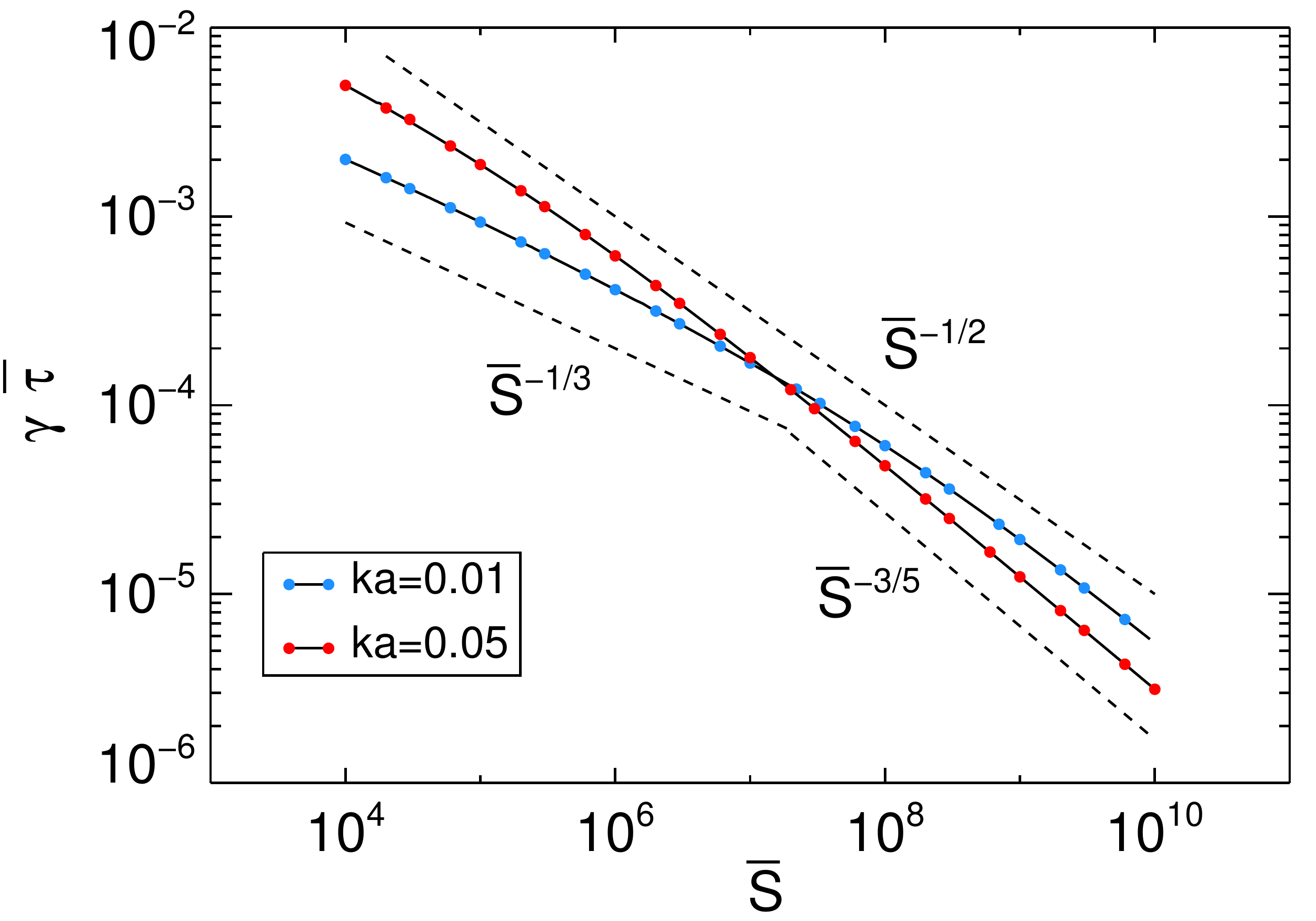}
\caption{Growth rate $\gamma\bar\tau$ vs. $\bar S$: transition from the small to the large $\Delta^{\prime}$ regime for two different wavevectors. Dashed lines represent the asymptotic scalings  of the large and small $\Delta^\prime$ regimes, $\gamma\bar\tau\sim\bar S^{-1/3}$ and  $\gamma\bar\tau\sim\bar S^{-3/5}$, respectively,   and of the fastest growing mode $\gamma\bar\tau\sim\bar S^{-1/2}$. The latter envelops the slope breaks occurring at the transition between small and large $\Delta^\prime$.}
\label{regimi}
\end{center}
\end{figure}

In Fig.~\ref{regimi} we show the transition from the small to the large $\Delta^\prime$ regime by plotting $\gamma\bar\tau$ as a function of $\bar S$ at two different wavevectors, $ka=0.01$ (light-blue dots) and $ka=0.05$ (red dots). The dashed lines correspond to the asymptotic scalings  of the growth rate given in eqs.~(\ref{small})--(\ref{fgm}). As can be seen, as the Lundquist number $\bar S$ increases, the growth rate for a given wave vector moves from large to small $\Delta^\prime$ regimes. The transition is marked by a break in the slope of $\gamma\bar\tau$ when the given wave vector corresponds to the fastest growing mode for a specific pair of $\{\gamma\bar\tau,\bar S\}_{k=k_m}$, so that the envelope of the break-points for all $k$s scales  as $\bar S^{-1/2}$, as expected.  

\section{Stability of thin current sheets and the ``ideal" tearing mode}
\label{ideal}

 In the traditional theory of tearing mode the current sheet aspect ratio is fixed and  its thickness $a$ is assumed to be macroscopic. 
On the other hand if a current sheet becomes thin enough, both the Alfv\'en time which normalizes the growth rate \emph{and} the Lundquist number become smaller and smaller. Therefore, even if the growth rate \emph{appears} to be small, it might actually be \emph{large} when  physically calculated in terms of macroscopic quantities. Indeed, as first seen by~\cite{bisk_1986}, the thin SP current sheet appeared to become unstable to fast reconnecting mode once a critical Lundquist number $S_c$ of order $10^4$ was passed.
It is therefore of interest to  consider generic current sheets whose aspect ratio scales with $S$ as $L/a\sim S^{\alpha}$, where now the scaling exponent~$\alpha$ ($\alpha=1/2$ for SP) is not specified~\citep{batta_2009,pucci}. For the sake of simplicity, we assume that the background magnetic field corresponds to a Harris current sheet. Note that current sheets scaling as $L/a\sim S^\alpha$ diffuse on a timescale $\tau_\eta$ that is $\tau_\eta\sim \taua\,S^{-1+2\alpha}$. Plasma flows are therefore a necessary part of SP sheet equilibrium, which otherwise would  diffuse in one  Alfv\'en time. Indeed, the SP configuration is based precisely on the requirement that convective transport of magnetic flux balances ohmic diffusion. Static equilibria can instead be constructed for thicker current sheets ($\alpha<1/2$), as they diffuse over a timescale much longer than the ideal one, $\tau_\eta/\taua\gg1$ for $S\gg1$. 

\begin{figure}
\begin{center}
\includegraphics[width=0.49\textwidth]{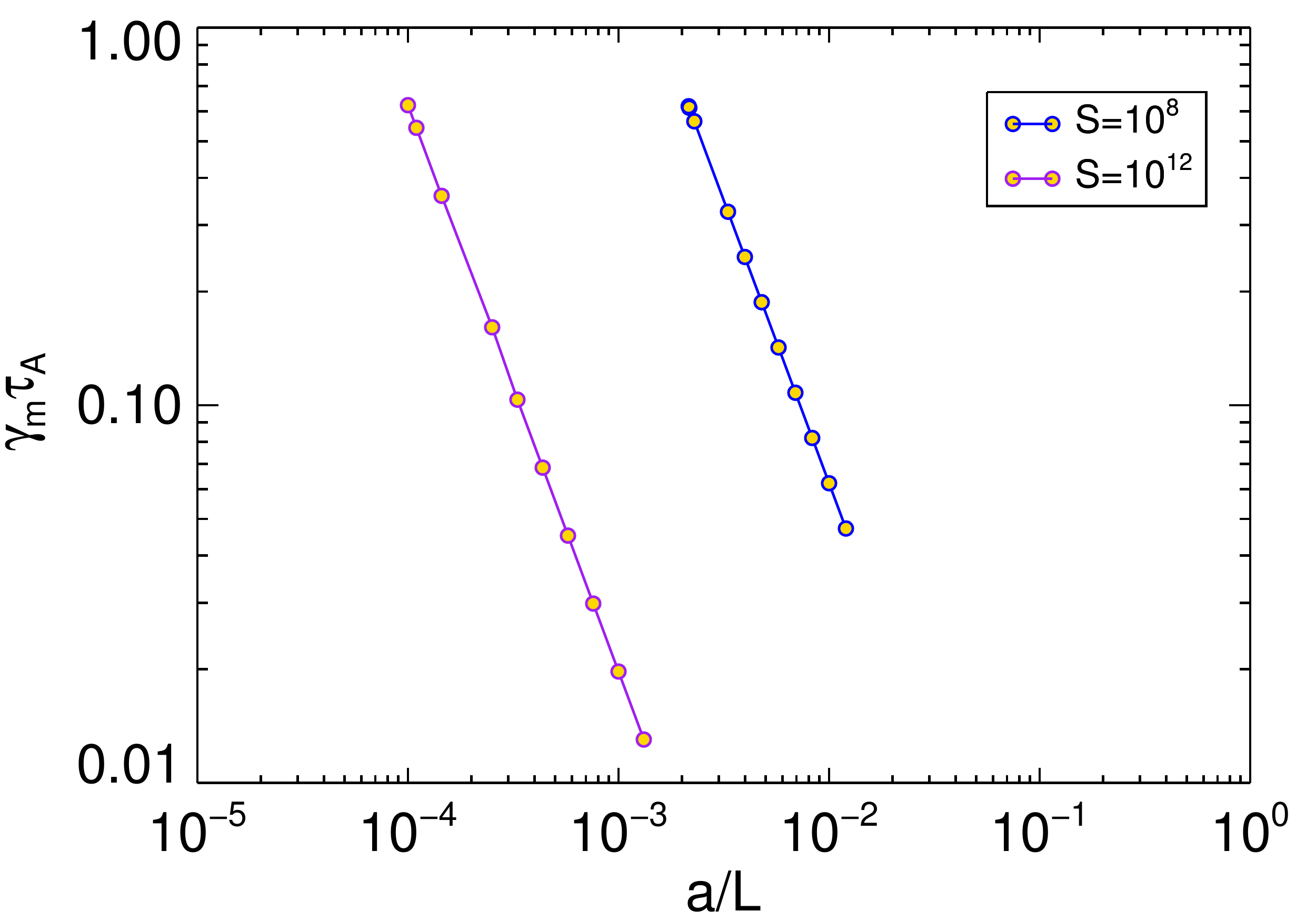}
\includegraphics[width=0.49\textwidth]{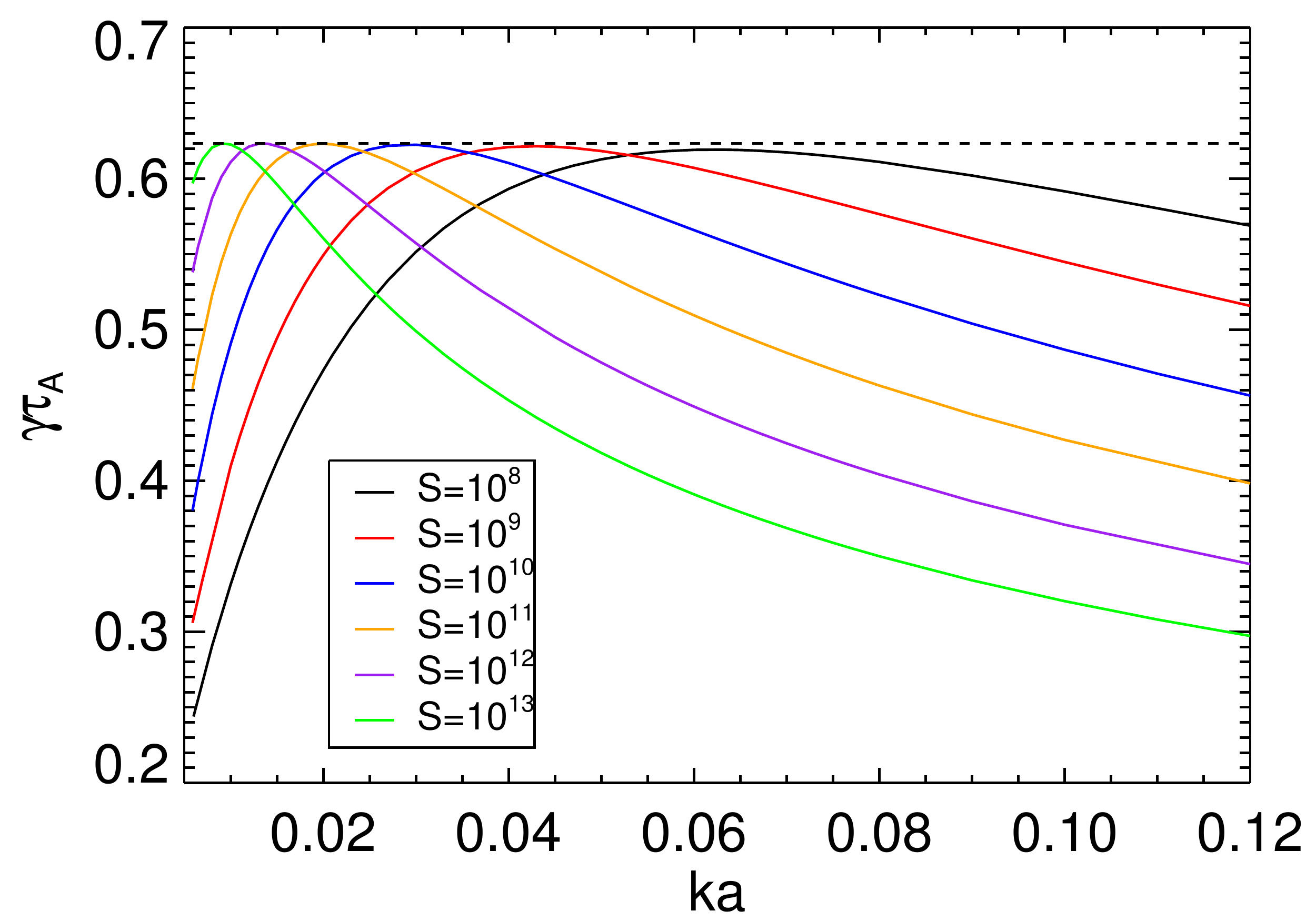}
\caption{Left:  growth rates normalized to the Alfv\'en time as a function of the inverse aspect ratio. Right: dispersion relations at $a/L=S^{-1/3}$ for different values of  $S$ (from~\cite{pucci}).}
\label{gam}
\end{center}
\end{figure}

The main idea underlying ``ideal" tearing  is the rescaling of the Lundquist number, which in the traditional tearing analysis is based on the thickness $a$ of the (macroscopic) current sheet equilibrium. If instead the length $L$ of the current sheet is considered as the macroscopic one, then the aspect ratio $L/a$ enters as a free parameter in the theory by introducing the renormalized quantities  $\taua=\bar\tau(L/a)$ and $S=\bar S(L/a)$. This leads to a maximum growth rate of the tearing instability which increases with the aspect ratio:
\begin{equation}
\gamma_m\taua\sim S^{-1/2}\left(\frac{L}{a}\right)^{3/2}\Rightarrow \gamma_m\taua\sim S^{-1/2+3\alpha/2}.
\label{GR}
\end{equation}
Fig.~\ref{gam}, left panel, shows the normalized maximum growth rate $\gamma_m\taua$ as a function of the inverse aspect ratio,  that confirms the theoretical scaling given by eq.~(\ref{GR}).

Eq.~(\ref{GR}) shows that current sheets having an aspect ratio that scales as a power $\alpha$ of the Lundquist number $\alpha>1/3$ are tearing-unstable with a maximum growth rate that diverges for $S\rightarrow\infty$. In particular, the growth rate of the plasmoid instability, $\gamma\taua\sim S^{1/4}$, is recovered for the scaling exponent of the SP sheet $\alpha=1/2$. This comes from the fact that in their original study  \cite{lou_2007} neglect the effects of the equilibrium flows, therefore reducing the calculation to a standard tearing mode boundary layer analysis (\citet{lou_2013} later found that flows seem to have negligible effect on the tearing mode within a SP sheet, though this result may be a consequence of the ordering assumptions).  Current sheets at  aspect ratios scaling with a power $\alpha<1/3$, on the contrary, are quasi-stable, i.e., $\gamma_m(\alpha<1/3)\rightarrow0$ for $S\rightarrow\infty$. When instead $\alpha=1/3$, current sheets are ``ideally" unstable in the sense that the corresponding maximum growth rate becomes  independent of $S$, and hence of order unity: as can be seen from Fig.~\ref{gam}, right panel, $\gamma_m\taua\rightarrow\gamma_i\taua\simeq0.63$ for $S\rightarrow\infty$ (where the index ``$i$" stands for ``ideal"). As such, the aspect ratio  $L/a\sim S^{1/3}$ provides an upper limit to current sheets that can naturally form in large Lundquist number plasmas, before they disrupt on the ideal timescale. Numerical simulations of a thinning current sheet, in which the thickness~$a$ is parameterized in time, show that indeed the current sheet disrupts in a few Alfv\'en times via the onset of ``ideal" tearing, when the critical thickness is approached from above~\citep{anna2},  as we will discuss in more detail in Section~\ref{trigger}.

To summarize, the ideally unstable current sheet has an inverse aspect ratio $a_i/L\sim S^{-1/3}$, with the wavevector and inner layer  thickness of the  fastest growing mode scaling with the Lundquist number  as
\begin{equation}
   k_iL \sim  S^{1/6},\quad\quad\frac{\delta_i}{L}\sim  S^{-1/2}.
\label{id}
\end{equation}

An interesting point to remark is that the so-called {\it inner diffusion} or {\it singular} layer of the ideally unstable current sheet has an aspect ratio which scales with the Lundquist number in the same way as the SP sheet,  $\delta_i/L\sim S^{-1/2}$. Yet, it is not a SP layer, because it does not correspond to a stationary equilibrium solution of the resistive MHD equations. The associated plasma flows  into the X-points and along the inner sheet itself, $\tilde u_{in}$ and $\tilde u_{out}$, are increasing exponentially in time together with the reconnected flux, and their ratio does not follow the SP scaling but rather  $\tilde u_{in}/\tilde u_{out}\sim S^{-1/3}$: in this accelerating growing mode the ratio of inflow to outflow velocity is larger. The latter scaling property can be easily verified by exploiting the incompressibility condition $\tilde u_{in}/\delta_i\sim \tilde u_{out}k_{i}$ and eq.~(\ref{id}). It is not a coincidence that the inner diffusion layer of the ideally tearing sheet scales like the SP, since  the  magnetic flux $\psi$ must now diffuse and reconnect on the ideal Alfv\'en timescale $\taua$ there, and the only length-scale at which magnetic field diffuses on the Alfv\'en time is precisely $\delta_i/L\sim S^{-1/2}$. Stated in another way, within the inner layer  we can approximate the diffusive term in Faraday's equation with $\eta\,\psi''\sim \eta\,\delta^{-2}\,\psi=(S\,\taua)^{-1}\,(L/\delta_i)^{2}\,\psi$, which must balance the growth rate $\gamma\psi\sim\taua^{-1}\psi$, yielding $\delta_i/L\sim S^{-1/2}$.

Although  ``ideal" tearing has been considered here in the simpler case of resistive MHD, it  provides a general framework for understanding under which conditions a fast tearing mode can  develop. ``Ideal" tearing  can be extended to include other physical effects which  impact unstable current sheets, some of which are discussed below.

\subsection{Effects of plasma flows on current sheet stability}
\label{sub_outf}

The tearing mode within thin current sheets has been considered assuming static equilibria, even though current sheets form together with plasma flows into (inflows, $u_{in}$) and along (ouflows, $u_{out}$) the sheet itself. As we have discussed, if $\alpha<1/2$ static equilibria can however be constructed that do not diffuse, or rather that diffuse over a timescale $\tau_\eta$ much longer than the ideal timescale. More precisely, inflows have little effect on the tearing mode if the diffusion rate within the inner reconnective layer, $u_{in}/\delta$, is negligible with respect to  the growth rate~\citep{dobrott}, which is indeed the case for the ideal mode.

While equilibrium inflows can  be neglected at large values of $S$, previous works~\citep{bulanov, bulanov2} showed  that the inhomogeneous outflow along the sheet has instead a stabilizing effect on the tearing mode: outflows may therefore induce the formation of thinner sheets having an inverse aspect ratio $a_i/L\sim S^{-\alpha_c}$, with $\alpha_c>1/3$. 

As discussed heuristically in~\cite{bisk_1986}, the tearing mode is stabilized when the  outflow rate $\Gamma_0\sim u_{out}/L\simeq \va/L$ exceeds a fraction of the growth rate,   $\Gamma_0> f \gamma_m$, and  this could explain the empirical critical Lundquist number $S_c\simeq10^4$ for the onset of  plasmoid instability within sheets having an aspect ratio scaling as SP.  The factor  $f\simeq 0.5$ takes into account that  growth rates deviate from their asymptotic values at low~$S$. It is  possible  to extend this argument in order to obtain the  scaling exponent $\alpha_c$ for ``ideal" tearing in current sheets with outflows~\citep{velli_unp,anna2}. Since $\gamma_m\taua=\gamma_i\taua S^{-1/2+3\alpha_c/2}$, where $\gamma_i\taua\simeq 0.63$, then the condition for an $S$-independent growth rate is given by $\Gamma_0=f\gamma_i S^{-1/2+3\alpha_c/2}$, that leads to
\begin{equation}
\alpha_c=\frac{2\log \mu+\log S}{3\log S}.
\label{outf}
\end{equation}
In eq.~(\ref{outf}),  $\mu=\Gamma_0/(f\gamma_i)$, and in particular the value $\mu=10$~\citep{bisk_1986} yields the observed $\alpha_c=1/2$ for $S = 10^4$, while, as expected,  $\displaystyle\lim_{S\to\infty}\alpha_c=1/3$. 

The impact of flows on the critical aspect ratio, expressed by eq.~(\ref{outf}), is consistent with recent numerical results~\citep{anna2}. Nevertheless, a more rigorous analysis of the effects of inflow-outflows on the stability of current sheets is still lacking, and a satisfactory explanation of the empirical stability threshold at low Lundquist numbers, $S\leq10^4$, and its possible dependence on initial/boundary conditions, remains to be given. Note that outflows should also impact the number of islands which develop in any given simulation close to the stability threshold.

\subsection{Impact of viscosity on the critical aspect ratio}
Viscosity is relevant in many astrophysical environments (e.g., the interstellar medium), in the laboratory,  and often in numerical simulations. The question therefore naturally arises as to how viscosity may impact the ``ideally"unstable current sheets. 

\begin{figure}
\begin{center}
\includegraphics[width=0.7\textwidth]{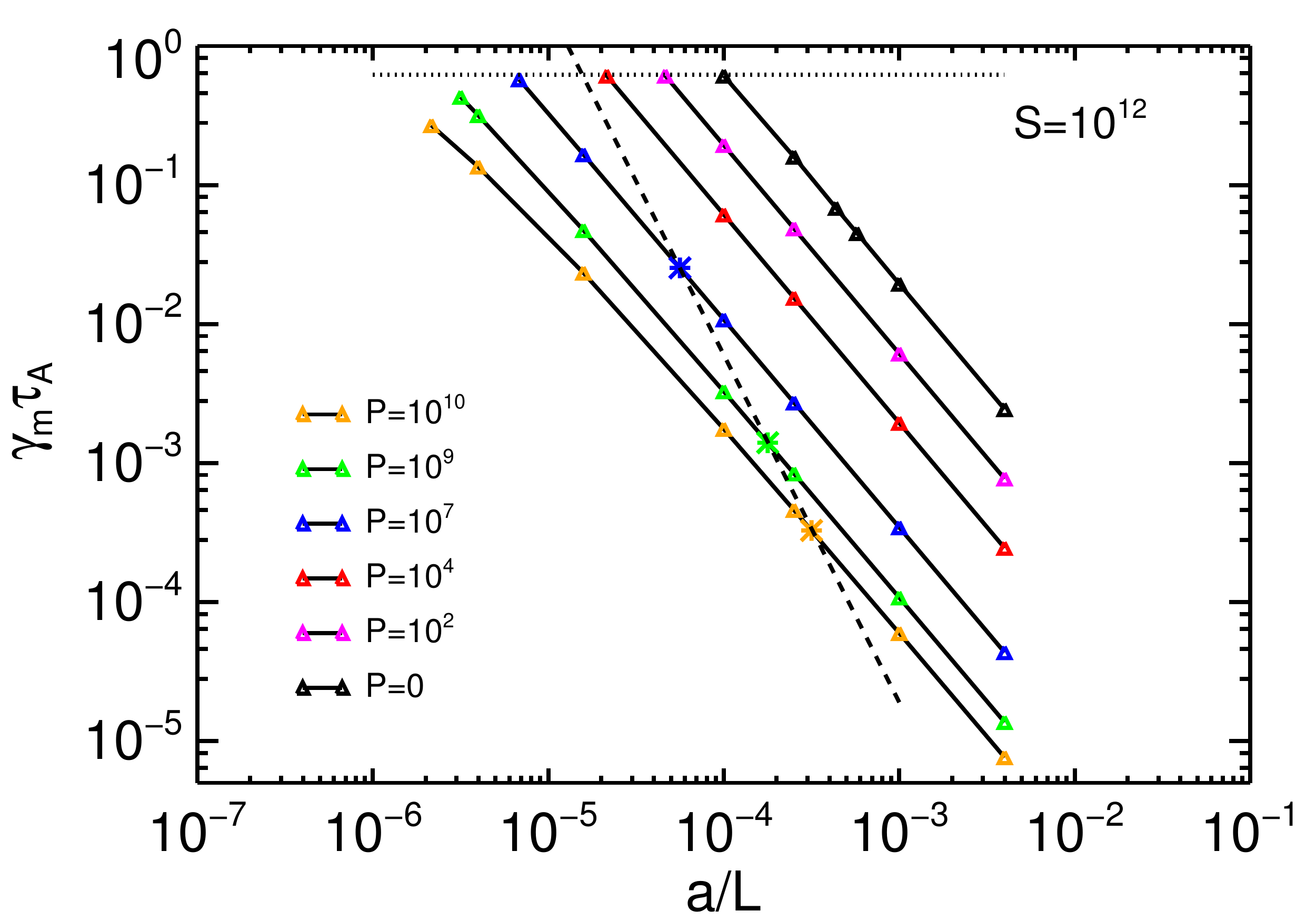}
\caption{Growth rate normalized to the Alfv\'en time vs. inverse aspect ratio at different Prandtl numbers. Stars correspond to the inverse aspect ratio of the viscous Sweet-Parker (from~\cite{anna1}).}
\label{visco}
\end{center}
\end{figure}

Effects of perpendicular viscosity on tearing modes in both the constant-$\psi$ and non constant-$\psi$ regimes have been addressed by many authors in the past, showing that viscosity reduces the tearing mode growth rate~\citep{porcelli_1987, bondeson, grasso_pop_2008, militello_pop_2011}. As a consequence, more elongated current sheets can form in a way similar, in some sense, to what happens in the presence of outflows. \citet{anna1} have shown that for large Prandtl numbers $P=\nu/\eta\gg1$ ($\nu$ is the  perpendicular kinematic  viscosity)   the fastest growing mode has a growth rate that scales with $S$, $P$ and the aspect ratio $L/a$ as
\begin{equation}
\gamma_m\taua\sim S^{-1/2}P^{-1/4}\left(\frac{L}{a}\right)^{3/2}\qquad\text{($P\gg1$),}
\label{viscomax}
\end{equation}
while the same scaling given by eq.~(\ref{GR}) holds for $P<1$. In Fig.~\ref{visco} we plot some values of the maximum growth rate as a function of the inverse aspect ratio at different Prandtl numbers,  for $S=10^{12}$. From eq.~(\ref{viscomax}) it is now possible to infer the scaling with $S$ and $P$ of the critical inverse aspect ratio, i.e., the one  corresponding to an ``ideal" growth rate. In this case, the ``ideally" unstable current sheet is thinner by a factor $P^{-1/6}$ than the one in a non viscous plasma, and is given by
\begin{equation}
\frac{a_i}{L}\sim S^{-1/3}P^{-1/6}\qquad\text{($P\gg1$)}.
\label{aivisco}
\end{equation}
We conclude by noting that the Sweet-Parker current sheet in the presence of viscosity  is instead  thicker with respect to the inviscid case, $a_{\text{\sc sp}}/L=S^{-1/2}(1+P)^{1/4}$~\citep{park}, whose inverse aspect ratio is indicated by the colored stars in Fig.~\ref{visco}. Therefore, for large Prandtl numbers,  $P\geq S^{2/5}$, the SP sheet can be quasi-stable to the tearing mode.  

\subsection{Collisionless reconnection in the ideal regime}
In low-collision regimes where resistivity is effectively negligible, the dominant effects violating the ideal Ohm's law are  electron inertia and/or  anisotropic electron pressure tensors. While both can allow magnetic reconnection, the latter usually requires a kinetic model~\citep{Cai,Scudder}, so  the electron skin depth $d_e\equiv c/\omega_{pe}\propto \sqrt{m_e}$ often appears as the only non-ideal term driving collisionless tearing modes when a fluid description of the plasma is adopted. In this case, the small parameter is not $S^{-1}$ but rather the normalized electron skin depth $d_e/L$. 

As a first step, we have considered the regime in which  magnetic reconnection is induced by electron inertia  in the strong guide field limit. The latter can be described by the Reduced MHD model in which the Hall term in Ohm's law can be neglected to the lowest order~\citep{strauss, Schep, delsarto_2006}. This description is suited, for instance, for tokamak and magnetically confined plasmas at low-$\beta$, where $\beta=8\pi P/B^2$ is the thermal to magnetic pressure ratio. It has been shown by~\cite{ideal_de} that in this regime the wavevector and growth rate of the fastest growing mode scale with the system parameters as 
\begin{equation}
k_ma\sim \frac{d_e}{a},\qquad\gamma_m\taua\sim \left(\frac{d_e}{L}\right)^2\left(\frac{L}{a}\right)^3,\qquad \frac{\delta_m}{a}\sim \frac{d_e}{a}.
\label{rmhd1}
 \end{equation}
From the scaling of the maximum growth rate $\gamma_m\taua$ in eq.~(\ref{rmhd1}) it follows that onset of ``ideal'' reconnection, that is when the growth rate becomes independent of the small non ideal parameter $d_e/L$, occurs once a critical aspect ratio 
\begin{equation}
\frac{a_i}{L}\sim \left(\frac {d_e}{L}\right)^{2/3} 
\label{rmhd2}
 \end{equation}
is reached. In the limit of strong guide field it is  possible to retain also some kinetic corrections related  to electron temperature anisotropies and ion Finite-Larmor-Radius (FLR) effects  in a simple way~\citep{Schep,Waelbroeck}, that can be included in the Reduced MHD model when approximated to their dominant gyrotropic contribution (we then speak of gyrofluid models). These models can  be used also at relatively large $\beta$ values~\citep{grasso2010}. Electron temperature effects arise at length-scales of the order of the ion sound Larmor radius $\rho_s=c_{s}/\Omega_i$, $c_{s}=\sqrt{T_e/m_i}$ being the ion sound speed and $\Omega_i$ the ion cyclotron frequency. These additional terms in the generalized Ohm's law can be seen as the result of an anisotropic electron pressure contribution~\citep{Schep}, or as the Hall term contribution when the diamagnetic drift is retained as a first order correction to the ${\bf E}\times{\bf B}$ electron drift (see also~\cite{ideal_de}). Ion FLR effects  introduce as a further length-scale the ion Larmor radius $\rho_i=v_{th}^i/\Omega_i$, where $v_{th}^{i}$ is the ion thermal speed. Depending on the way ion FLR are approximated from kinetic theory, different gyrofluid models can be obtained, all yielding the same dispersion relation for finite temperature tearing instabilities when  $\rho_\tau^2\equiv\rho_s^2+\rho_i^2\gg d_e^2$~\citep{peg86,porcelli91, otta95}. An explicit expression for the fastest growing mode valid in this regime can be found in~\cite{Comisso}. For a Harris sheet, when rescaled from the sheet thickness $a$ to the length $L$, the wave-number and growth rate of the fastest growing mode are therefore the following~\citep{ideal_de},
\begin{equation}
k_ma\sim \frac{d_e}{L}\left(\frac{\rho_\tau}{d_e}\right)^{1/3}\frac{L}{a},\qquad \gamma_m\taua\sim \left(\frac{d_e}{L}\right)^2\frac{\rho_\tau}{d_e}\left(\frac{L}{a}\right)^{3}\qquad\text{($\rho_\tau\gg d_e$)}.
 \label{kin1}
\end{equation}
In this regime, the critical inverse aspect ratio scales with both $d_e/L$ and $\rho_\tau/d_e$ as
\begin{equation}
 \frac{a_i}{L}\sim\left(\frac{d_e}{L}\right)^{2/3}\left(\frac{\rho_\tau}{d_e}\right)^{1/3}\qquad\text{($\rho_\tau\gg d_e$)}.
 \label{kin2}
\end{equation}

In conclusion, kinetic effects increase the linear growth rate by a factor $\rho_\tau/d_e$ with respect to the low-$\beta$ case, as can be seen by comparing the growth rate in eq.~(\ref{rmhd1}) with the one including finite temperature effects in eq.~(\ref{kin1}). As a consequence, the critical inverse aspect ratio turns out to be thicker in the latter case by a factor scaling as $(\rho_\tau/d_e)^{1/3}$ (cfr. eq.~(\ref{rmhd2}) and eq.~(\ref{kin2})).
 
%%%%%%%%%%%%%%%%%
\section{The trigger problem}
\label{trigger}

Observations show that many explosive phenomena usually display an initial phase (of long but uncertain duration) during which energy is transferred to and accumulated in the system in question, followed by an abrupt destabilization process expressed by an impulsive phase, often attributed to the onset, or trigger, of magnetic reconnection. During the impulsive phase energy is suddenly released on a fast (ideal) timescale (of the order of a few $\taua$) in the form of heat, kinetic energy and populations of accelerated particles. The impulsive phase is followed by main and  recovery phases, whereby energy released becomes more gradual and  the system relaxes towards a quiet configuration~\citep{aka,wesson_1986,Fletcher}. Solar flares provide one of the most spectacular  examples in this sense: a  prominence can stand for weeks in the solar corona until it erupts by releasing  a huge amount of energy (the flare), of about $10^{30}-10^{32}$~erg, in a few tens of minutes or hours, depending on their size. The flare displays an impulsive phase, that can be seen as sudden intensity enhancements in different wavelengths, especially in the hard X-ray spectrum, lasting usually no more than a few minutes for the larger events~\citep{Ellison,kane74,ajello}.  

In the weakly collisional plasmas found in many astrophysical environments, or in fusion devices, bulk plasma motions tend to form thin current sheets spontaneously in localized regions~\citep{syro, bisk_1993}. There, the intense currents are limited only by the extremely small resistivity or by other effects, such as electron inertia, until they ultimately relax once reconnection is enabled inside these thin boundary layers. The problem of understanding in which way thin current sheets form is a complex and rich one, both theoretically and numerically. We therefore do not go into details, but some recent studies about thin current sheet formation for configurations relevant to magnetosphere dynamics and the solar corona can be found for instance in \cite{birn_2002,hsieh_2015,titov_2003,aulanier_2005, rappazzo}. 

Whatever the mechanism might be,  one can always imagine that, during current sheet formation, the system evolves through a sequence of similar configurations to which the linear stability analysis can be applied in order to investigate its properties. As shown in Section~\ref{ideal},  the growth rate of the reconnecting tearing mode instability exhibits a strong dependence on the current sheet aspect ratio, for example in resistive MHD $\gamma\taua\propto (L/a)^{3/2}$ (cf. eq.~(\ref{GR})). This, together with the large values of the Lundquist number, or $L/d_e$, suggests that the formation of small scales all the way down to the critical thickness would naturally lead to the  trigger of fast reconnection.  

\subsection{Scenario} 
In the following, we will assume a current profile that thins exponentially in time. Exponentially thinning current sheets are indeed of particular importance, as they are commonly observed in simulations of solar and  stellar coronal heating~\citep{rappazzo}, as self-similar solutions of  X-point collapse~\citep{sulem}, as well as in in-situ measurements  during the growth phase of substorms in the Earth's magnetotail~\citep{sanny_1994}, to give some examples. 

 We therefore assume  that the ambient plasma and magnetic fields evolve over a timescale $\tau_c$ which is determined by ideal plasma motions and hence does not depend on $S$.  Consider the simpler case of resistive (incompressible) MHD, and take a Harris current sheet of inverse aspect ratio $a/L$ shrinking on a timescale~$\tau_c\sim a\, (da/dt)^{-1}\gtrsim\taua$. The parameter $\tau_c$ thus  represents schematically  the coupling between the local dynamics within the current sheet and some external process inducing the thinning itself. Within this framework, we extend the tearing mode analysis to systems that  are parameterized in time through the evolving aspect ratio~\citep{anna2}. 

  An investigation of the  disruption of a forming current sheet germane to the present discussion has been recently published in ~\cite{uzdensky_2014}. Since the two scenarios differ somewhat, let us briefly comment on the differences here, though we refer the reader to that paper for more details. Our study focuses on the linear stage of the tearing instability to provide a possible explanation for the onset of fast reconnection, using a WKB approach to describe the growth of unstable modes during the dynamical collapse of the sheet. In this way, we take into account  that unstable modes with a growth rate exceeding $\tau_c^{-1}$ can in principle compete in disrupting the current sheet. \citet{uzdensky_2014} on the other hand consider arbitrary timescales for driving the sheet collapse, and start from the assumption that the linear stage of a mode ends once its growth rate exceeds the driving rate. They then estimate that the first mode to meet this condition will also be the one dominating the  subsequent nonlinear evolution. They therefore examine the nonlinear regime to describe the disruption of the current sheet, which in their scenario may occur also on timescales scaling as a positive power of $S$. In this sense, they do not impose restrictions on mode growth times with respect to dissipative coefficients. 
Our approach on the other hand focuses precisely on the latter aspect, our framework being one in which things will occur on the fastest possible times compatible with the dynamical system in question, and therefore, in analogy for example with turbulence,  we expect the fastest growing modes not to scale with the dissipative coefficients. So in some sense the approach of \citet{uzdensky_2014} is more general than ours, but does not investigate in detail whether and under which conditions the transition to fast reconnection (the trigger) might occur. Other recent interesting analyses of a similar conceptual reconnection problem, the Taylor's problem, have been provided by \citet{comisso15} and \citet{vek15}. They analyze the possible time evolution of an initial macroscopic, stable current layer which is subject to   a finite amplitude perturbation at its boundaries, and recognize the role sufficiently fast tearing modes may play in disrupting the current sheet. We do not discuss that problem here, though it would be interesting to study how in the asymptotic limit of large $S$ the ``ideal" tearing framework affects those models. Let us focus now on our scenario of an exponentially thinning sheet.

On the basis of the rescaling argument, the growth rates in the small and large $\Delta^\prime$ regimes discussed in Section~\ref{tearing} can now be generalized to arbitrary aspect ratios. By taking $\Delta^\prime a\sim[(ka)^{-1}-ka]$ for a Harris sheet, these are  given, respectively, by
\begin{equation}
\gamma\taua\sim S^{-3/5}\left(\frac{a}{L}\right)^{-8/5}\left(  ka \right)^{-2/5}[1-(ka)^2]^{4/5},\qquad \gamma\taua\sim S^{-1/3}\left(\frac{a}{L}\right)^{-2}\left(  ka \right)^{2/3}.
\label{small1}
\end{equation}
 The fastest growing mode, as already shown in Section~\ref{ideal},  has the growth rate scaling~as 
\begin{equation}
\gamma_m\taua\sim S^{-1/2}\left(\frac{a}{L}\right)^{-3/2},
\label{fgm1b}
\end{equation}
whereas the wave vector and thickness of inner diffusion layer scale, respectively, as
\begin{equation}
k_ma\sim S^{-1/4}\left(\frac{a}{L}\right)^{-1/4},\qquad \frac{\delta_m}{a}\sim S^{-1/4}\left(\frac{a}{L}\right)^{-1/4}.
\label{fgm1}
\end{equation}
\begin{figure}
\begin{center}
\includegraphics[width=0.47\textwidth]{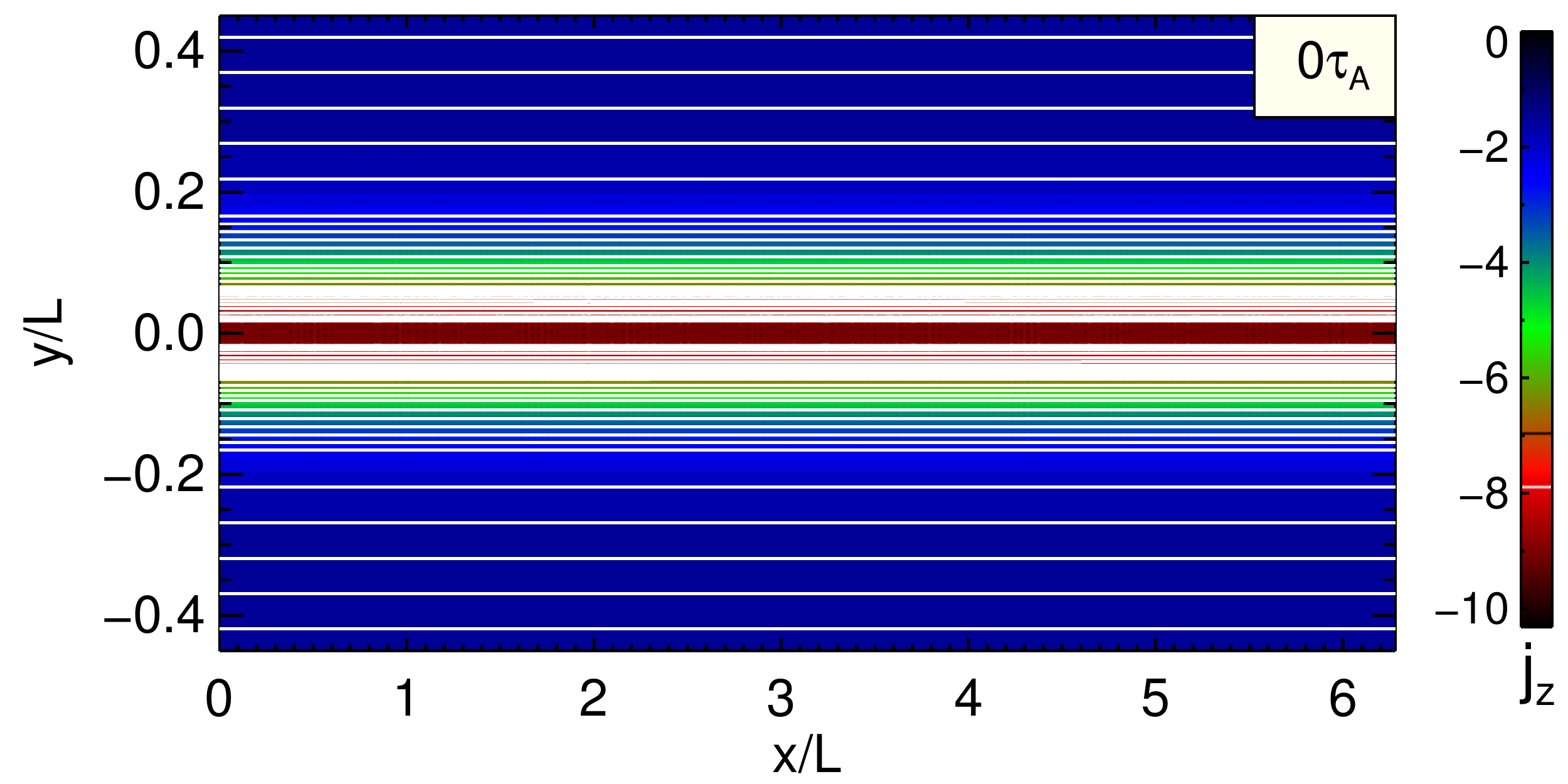}
\includegraphics[width=0.47\textwidth]{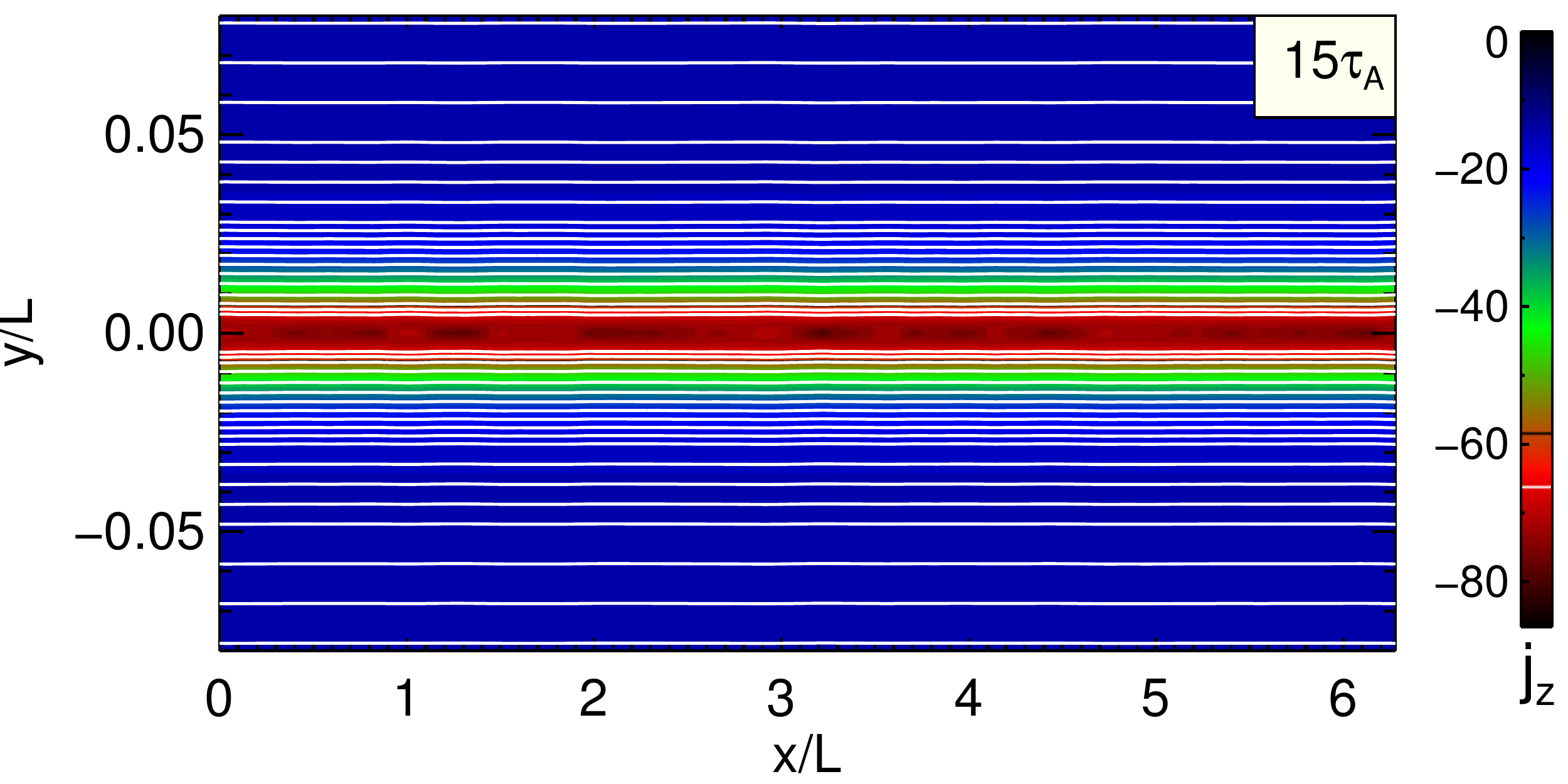}
\includegraphics[width=0.47\textwidth]{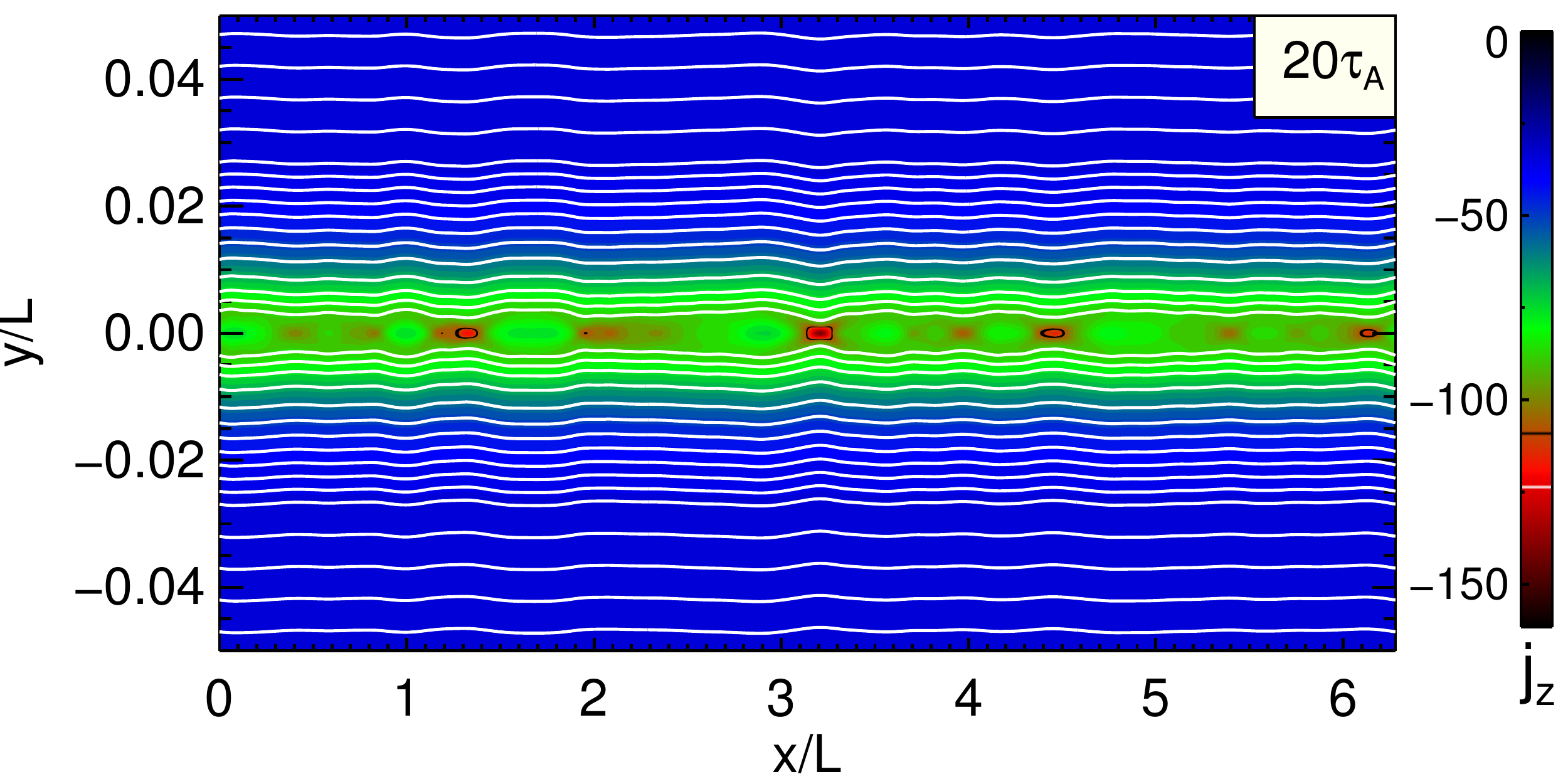}
\includegraphics[width=0.47\textwidth]{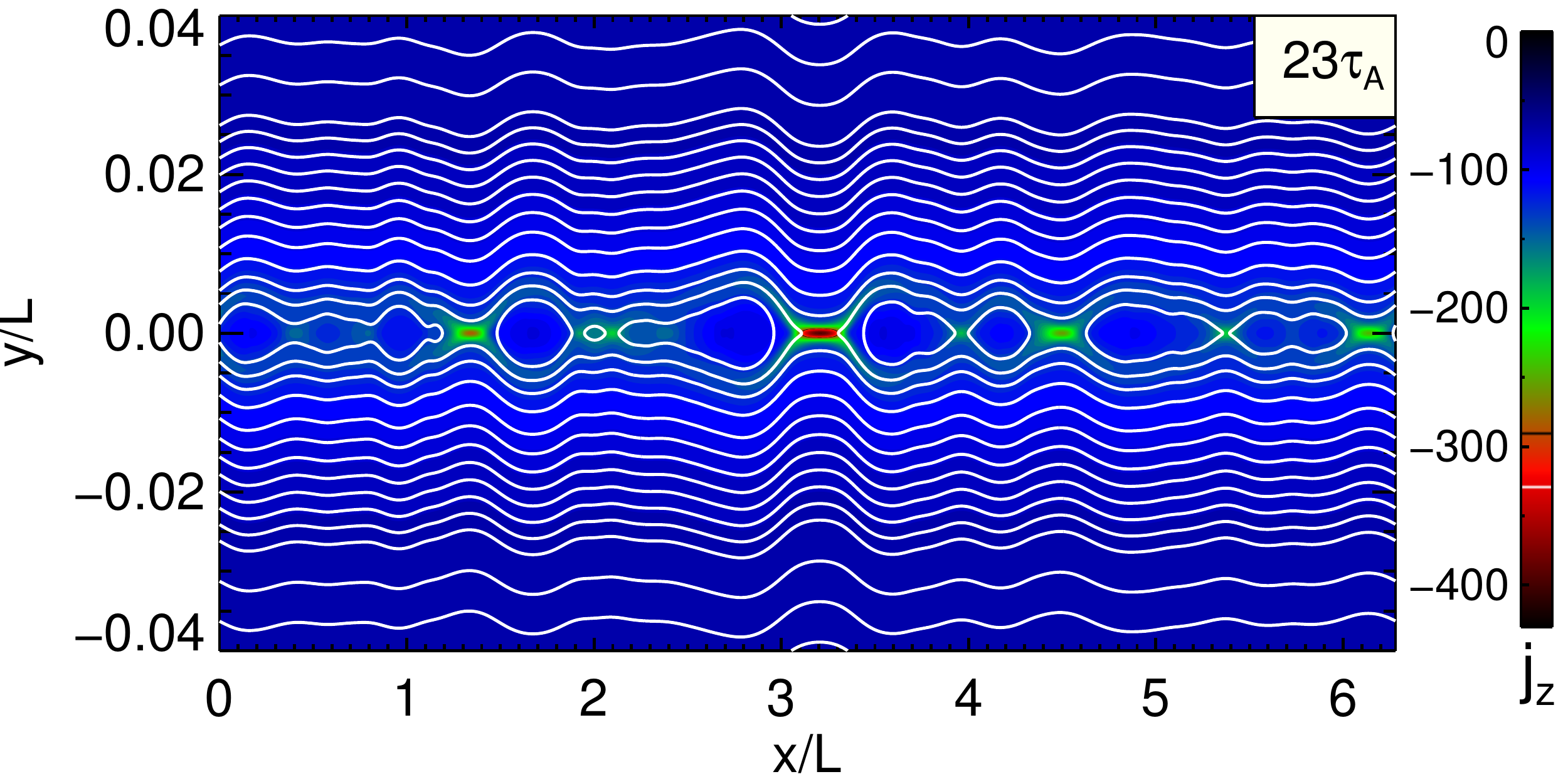}
\includegraphics[width=0.47\textwidth]{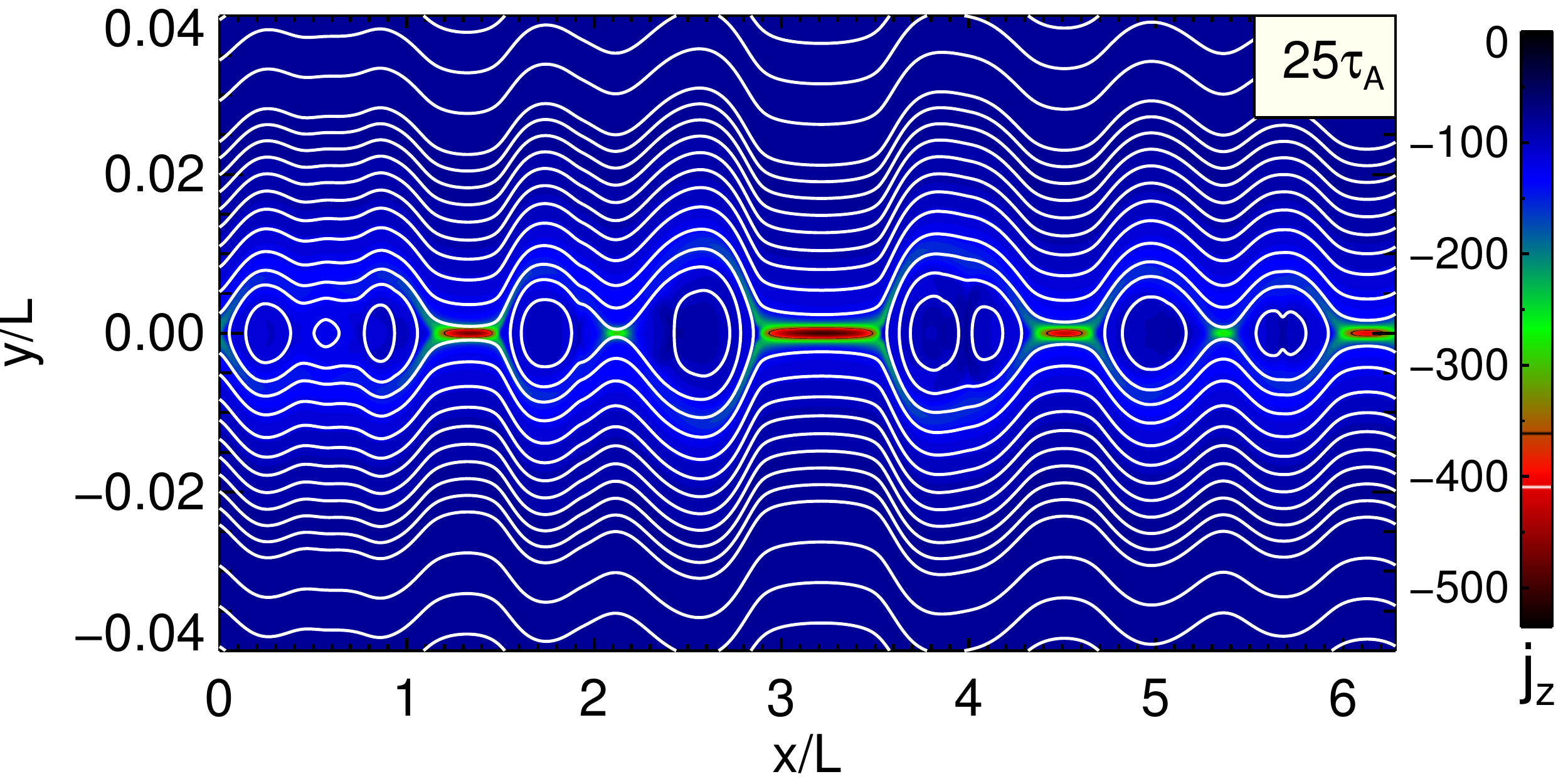}
\includegraphics[width=0.47\textwidth]{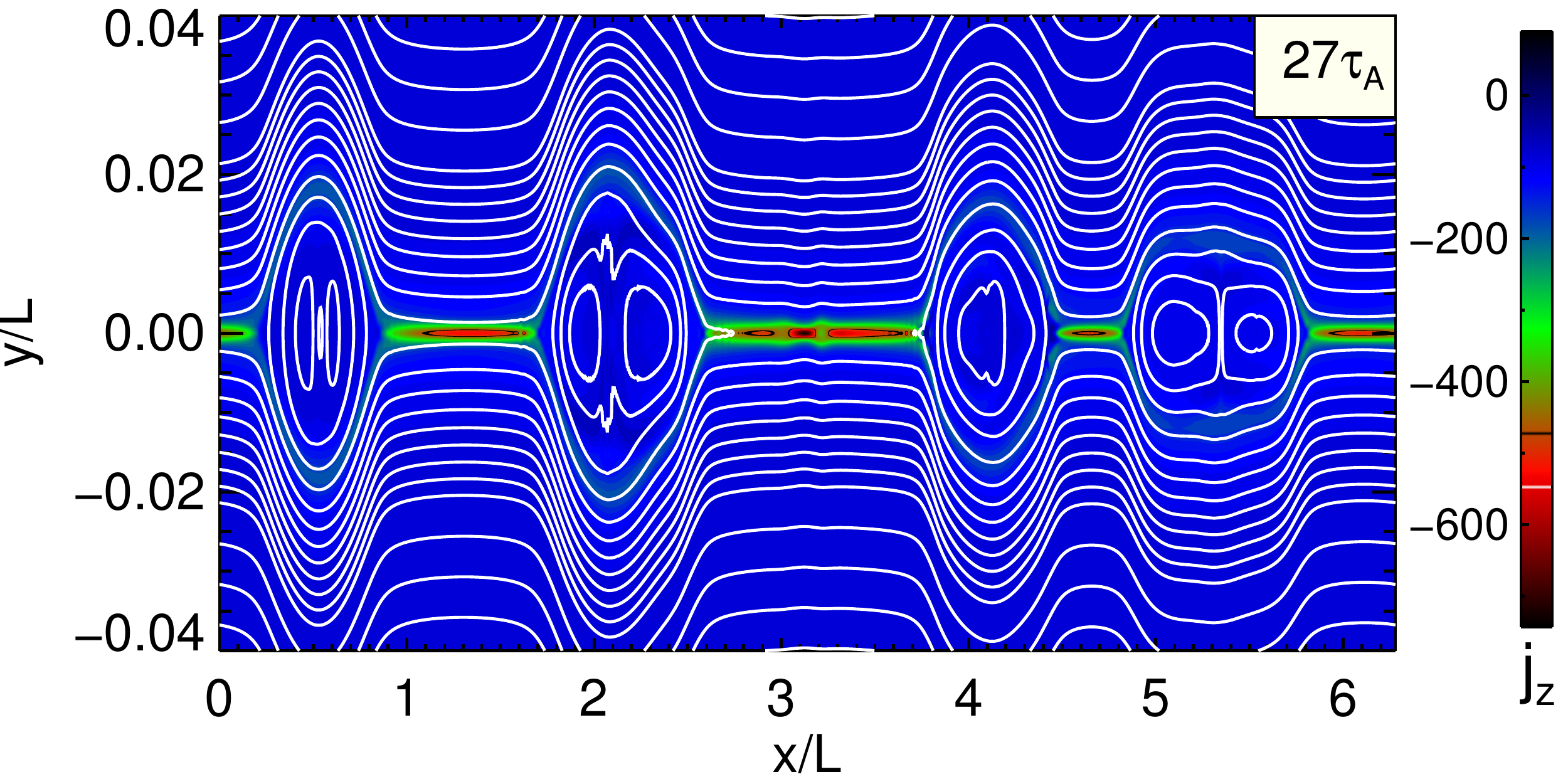}
\caption{Temporal evolution of  a thinning Harris current sheet. The out-of-plane current density $j_z$ is color coded and white lines represent magnetic field lines.}
\label{coll}
\end{center}
\end{figure}

As the thinning proceeds in time, modes with increasing wavevector $k$ are progressively destabilized, owing to the traditional instability condition $ka(t)<1$ in a Harris sheet. In this way, each newly destabilized mode lies in the small $\Delta^\prime$ regime and it  transitions toward the large $\Delta^\prime$ one as time elapses, by crossing at some point  the fastest growing mode. This can be understood also by looking back at the plot in Fig.~\ref{regimi}: since $\bar S\equiv S (a/L)$,  if  $a/L$ decreases at fixed $S$, then $\bar S$ decreases accordingly, so that a given mode shifts from the right   to left side of the plot. On the other hand, only those modes fitting within the current sheet, that therefore satisfy $kL\geq2\pi$, can grow. This  implies that at the beginning, when $a\lesssim L$, there are only unstable modes in the small $\Delta^\prime$ regime. This initial stage in turn ends when the smallest unstable mode, which has a wavelength of the order of the current sheet length $L$, $k\sim2\pi/L$, coincides with  $k_m$~\citep{uzdensky_2014}. This happens when $S^{-1/4}(a/L)^{-1/4}\simeq2\pi (a/L)$, or, in other words, when  $a/L\lesssim 0.2 S^{-1/5}$. At that point, the fastest growing mode will be always unstable, and hence it will dominate over both the small and the large $\Delta^\prime$ modes. The initial stage evolving in the small $\Delta^\prime$ regime however  does not contribute significantly to the growth of perturbations so one can consider only the regime given by the fastest growing mode, expressed by eqs.~(\ref{fgm1b})--(\ref{fgm1}), to describe the transition to fast reconnection. Indeed, the small $\Delta^\prime$ (as well as the large $\Delta^\prime$) regime has a growth rate that, even if the current sheet is thinning, goes to zero for $S\rightarrow\infty$ (according to eq.~(\ref{small1})). As a consequence, modes in that regime cannot grow during a finite interval of time of order of $\tau_c$. In other words, the small $\Delta^\prime$ regime corresponds to {\it slow} reconnection and it cannot in general provide the transition to fast reconnection in quasi-ideal plasmas.
Assuming a collapse of the form $a(t)=\exp(-t/\tau_c)$, which can  be easily generalized to include an exponential increase of the length $L$,   then the amplitude of the reconnecting magnetic flux $\psi$ is given~by 
\begin{equation}
 \psi(t)=\psi(0)\exp{\left(\int^t_0\gamma(t^\prime)dt^\prime\right)},
\label{int}
\end{equation}
where, by neglecting the initial small $\Delta^\prime$ stage (current sheets having $a/L>S^{-1/5}$), the integrand is  given~by
\begin{equation}
\gamma(t)=\gamma_m(t)=\frac{1}{\taua}\,0.63\,S^{-1/2}\,\left[\exp\left(\frac{t}{\tau_c}\right)\right]^{3/2}.
\label{integrand}
\end{equation}
Expressions~(\ref{int})--(\ref{integrand}) yield 
\begin{equation}
 \psi(t)=\psi(0)\exp\left[   \frac{2\tau_c}{3} 0.63\, S^{-1/2}\left( e^{3t/(2\tau_c)}-1 \right) \right],
\end{equation}
that can be approximated as follows,
\begin{equation}
 \psi(t)\simeq \psi(0)e^{\gamma_m(0)t},\quad\text{if $t\ll \tau_c\,2/3$},
\label{int1}
\end{equation}
\begin{equation}
 \psi(t)\simeq \psi(0)e^{\gamma_m(t)2\tau_c/3},\quad\text{if $t> \tau_c\,2/3$}.
\label{int2}
\end{equation}
\begin{figure}
\begin{center}
\caption{Temporal evolution of the current sheet thickness (upper panel) and of the amplitude of some Fourier modes of the flux function $\psi$ at the neutral line (lower panel) of the simulation shown in Fig.~\ref{coll}. The dashed line corresponds to the integral given in eq.~(\ref{int}) where the functional form $a(t)$ adopted in the simulation (see text) has been used~(from \citet{anna2}).}
\label{modes_R4}
\end{center}
\end{figure}

The onset of the tearing instability roughly  takes place when unstable modes have time to develop, hence at about the time $\tau_*$ such that $\gamma_m(\tau_*)\tau_c\simeq1$, as can be seen also from eq.~(\ref{int2}). In turn, the growth rate rapidly increases while approaching the critical thickness $a_i/L\sim S^{-1/3}$ from above, so one can easily convince himself  that a fast current sheet collapse naturally drives an explosive transition from a quasi-stable state -- growth rate depending on a negative power of $S$ -- to an ideally unstable one. 

 We have tested such scenario with fully nonlinear MHD simulations in which a collapse of the current sheet is imposed  {\it a priori} with $\tau_c$ in the range $1-10\, \taua$~\citep{anna2}. Fig.~\ref{coll} shows an example of the temporal evolution of an exponentially thinning Harris sheet initially perturbed by a random noise of fluctuations.  { The Lundquist number in these simulations is set to $S=10^6$  and the Prandtl number  to $P=1$, so that the scalings  of the non-viscous case are not  modified significantly. } The in-plane magnetic field is ${\bf B}=B_0\tanh[y/a(t)]{\bf \hat x}$ and, for numerical convenience, a thinning of the form $a(t)=a_0\exp(-t/\tau_c)+a_\infty(1-\exp(-t/\tau_c))$ has been chosen. For the case shown in Fig.~\ref{coll} we have fixed $a_\infty/a_0=0.1$, with $a_0=0.1\,L$, and $\tau_c=4\,\taua$.  In Fig.~\ref{modes_R4} we show the temporal evolution of $a(t)$ (upper panel) and of some unstable Fourier modes of $\psi$ at the neutral line $y=0$ (lower panel) for the same simulation. The simulation  illustrates that during the linear stage of the instability (from $t\simeq11$ to $t\simeq 22$ $\taua$) the magnetic field  rapidly reconnects in a few Alfv\'en times when the critical thickness $a_i/L\sim S^{-1/3}\simeq0.01$ is approached from above, and that  the current breaks-up into a number of magnetic islands $N_i$ that scales as $N_i\sim S^{1/6}\simeq 10$.    As can be seen from Fig.~\ref{coll}, the following nonlinear stage is characterized by the competition between coalescence of magnetic islands (this can be seen also by looking at the increasing magnitude of long wavelength modes, $kL=5$ and $kL=1$, in Fig.~\ref{modes_R4}) and X-point collapse, {in a way qualitatively similar to that discussed in~\citet{malara_1992}}. We defer more detailed discussions about the nonlinear evolution of an ``ideally" reconnecting current sheet to Section~\ref{nonlinear}. 

To summarize, in the scenario proposed above the mechanism driving the collapse  (that can be mapped in the developmental phase, or the storage phase) is not specified. The ambient current sheet thins on a timescale $\tau_c$ which is independent from $S$ and instability is let to freely evolve from an initial slow phase of reconnection, when the thickness is larger than critical, to an ideally fast one while approaching the critical inverse aspect ratio $a_i/L\sim S^{-1/3}$. Numerical simulations of a thinning current sheet with $\tau_c\simeq1-10\,\taua$ show that when approaching the critical inverse aspect ratio from above magnetic islands significantly grow from the initial noise up to the thickness of the inner layer of the reconnecting current sheet itself and beyond, during the nonlinear stage, on a timescale of the order of a few $\taua$. In the model that we have chosen of an exponential collapse the time needed to form the ideally unstable current sheet is about $\tau_i\simeq(\tau_c/3)\ln(S)$, i.e., it depends very weakly (logarithmically) on $S$ and allows for a two timescale dynamics. For instance,  if $\tau_c=1-10\,\taua$ and $S=10^{12}-10^{14}$ then $\tau_i\simeq 10-100\,\taua$ thus one or two orders of magnitude larger than the time expected for the onset and development of ``ideal" reconnection. 

Nonlinearities become important when the half-width $w$ of magnetic islands, given approximately by   $w\simeq 2\,a \sqrt{\psi/(B_0\,a)}$ {(see, e.g.,~\citet{bisk_2000} pp. 82--83)},  is of the order of the half-thickness of the inner diffusion region of the reconnecting sheet $\delta_m$,  right equation~(\ref{fgm1}). This condition is met when the perturbation has an amplitude scaling as $\psi_{nl}/(B_0L)\sim 0.25S^{-1/2}(a/L)^{1/2}$. For a critical current sheet $a_i/L\sim S^{-1/3}$ this estimate yields a nonlinear amplitude $\psi_{nl,i}/(B_0L)\sim 0.25\,S^{-2/3}$, and even smaller  for thinner sheets. For example, the Sweet-Parker with $a_{\text {\sc sp}}/L\sim S^{-1/2}$ has a nonlinear amplitude of about $\psi_{nl,\text{\sc sp}}/(B_0L)\sim 0.25\,S^{-3/4}$.  In solar active regions $B_0\sim100$~G,  $L\sim10^4$~km, and $S\simeq10^{12}$, yielding  $\psi_{nl}\sim 0.25 \times10^{-9} \times (10^4 \times 100) =0.25 \times 10^{-3}$ G km with background magnetic flux of about $\psi_0\sim (B_0L)S^{-1/3}\sim100$~G km, for $a/L\simeq S^{-1/3}$. Therefore, the initial perturbation should be about $0.25\times 10^{-5}$ times smaller than the background. The duration of the linear phase depends on both  the initial noise level and the growth rate of the instability (and it becomes shorter and shorter with $S$ for  sheets thinner than critical). It is certainly clear that once the ``ideal" instability limit is approached the duration of the linear phase may be extremely limited. Nonetheless  if the linear theory is thought of as providing a trigger for faster, nonlinear dynamics, as we will discuss in the next section, the ``ideally" unstable aspect ratio remains an upper limit for laminar current sheets. It is reasonable to expect that an initial nonlinear noise  would lead to the disruption of the current sheet, and to a turbulent regime, on ideal timescales via the growth of resonant (most unstable) modes. In this regard a detailed study requires numerical simulations at different nonlinear noise levels in some sense generalizing the paper by~\citet{matt}.

\begin{figure}
\begin{center}
\includegraphics[width=\textwidth]{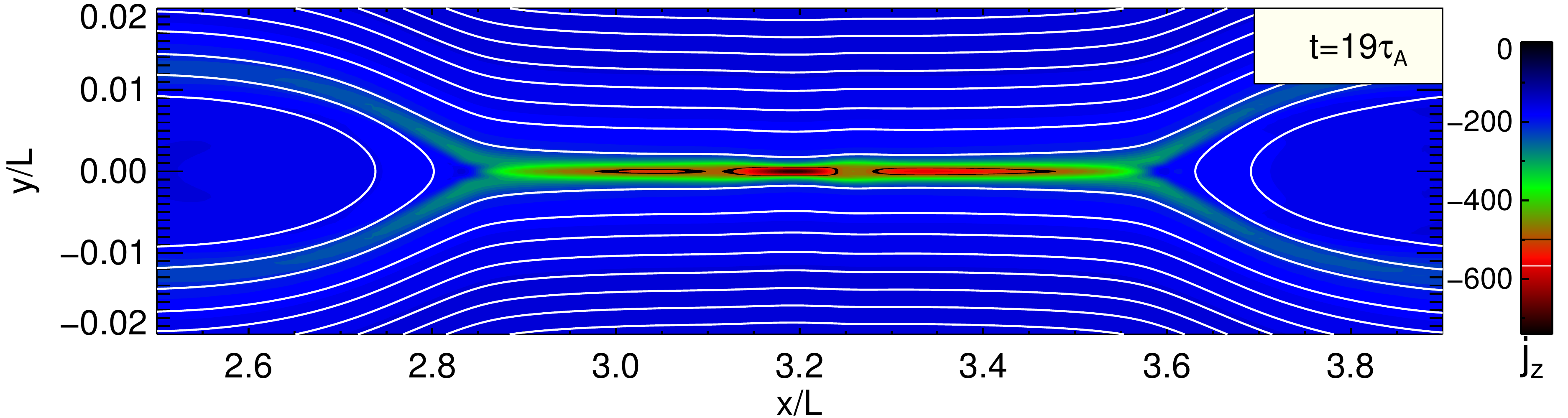}
\caption{Blow-up of the collapsing current sheet during the nonlinear stage of the primary ideal tearing instability (from~\citet{anna2}).}
\label{bolle}
\end{center}
\end{figure}

\section{Recursive reconnection: the ``fractal" reconnection model revised}
\label{nonlinear}

X-points arising from a  tearing instability at the largest wavelengths, i.e., far from the small $\Delta^\prime$ regime,  collapse  into an elongated current sheet during the early nonlinear stage~\citep{wael,jemella}. In this respect,  nonlinearities of the tearing mode provide a self-consistent mechanism for current sheet formation to which the scenario discussed in Section~\ref{trigger} can in turn be applied, on the basis of similarity and rescaling of length and timescales. This simple idea led  \citet{shibata} to  propose a phenomenological  ``fractal reconnection" model for flares, that we revisit here in light of the ``ideal" tearing scenario.

In general, it is observed  that if the aspect ratio of the secondary current sheet becomes large enough, then the latter becomes unstable to secondary tearing  generating secondary  plasmoids~{\citep{malara_1992,Lou_2005}}. This can be seen in the last panel of Fig.~\ref{coll} and in a blow-up of one of the secondary current sheets, displayed in Fig.~\ref{bolle}, the latter obtained from a similar simulation with {$S=10^6$, $P=1$}, $\tau_c=\taua$, $a_\infty/a_0=0.1$ and $a_0=0.1\, L$. A number of past numerical and theoretical studies have addressed the problem of the transition from the laminar reconnection, proper of the early nonlinear stage, to the subsequent  highly unsteady one, characterized by intermittent generation of plasmoids within the sheet itself, bearing faster average reconnection rates~\citep{lapenta_2008,batta_2009}.  {The Sweet-Parker paradigm has been constantly invoked in the interpretation of the nonlinear evolution -- both to explain the instability onset within the laminar current sheet, and to model the following fully nonlinear plasmoid-dominated stage~\citep{Lou_2005,batta_2009,cassak_2009,dau_2009,huang_2010,uzdensky_2010,ali_2014}. However, the onset of fast reconnection may take place at smaller aspect ratios for a given Lundquist number: in particular, ``ideal" tearing predicts a different instability criterion and different scaling laws for the dependence on (macroscopic) quantities $L$ and~$S$,  which, when the possible effects of flows are taken into account, may provide a better guide for inspecting the nonlinear evolution. }     

\begin{figure}
\begin{center}
\includegraphics[width=0.49\textwidth]{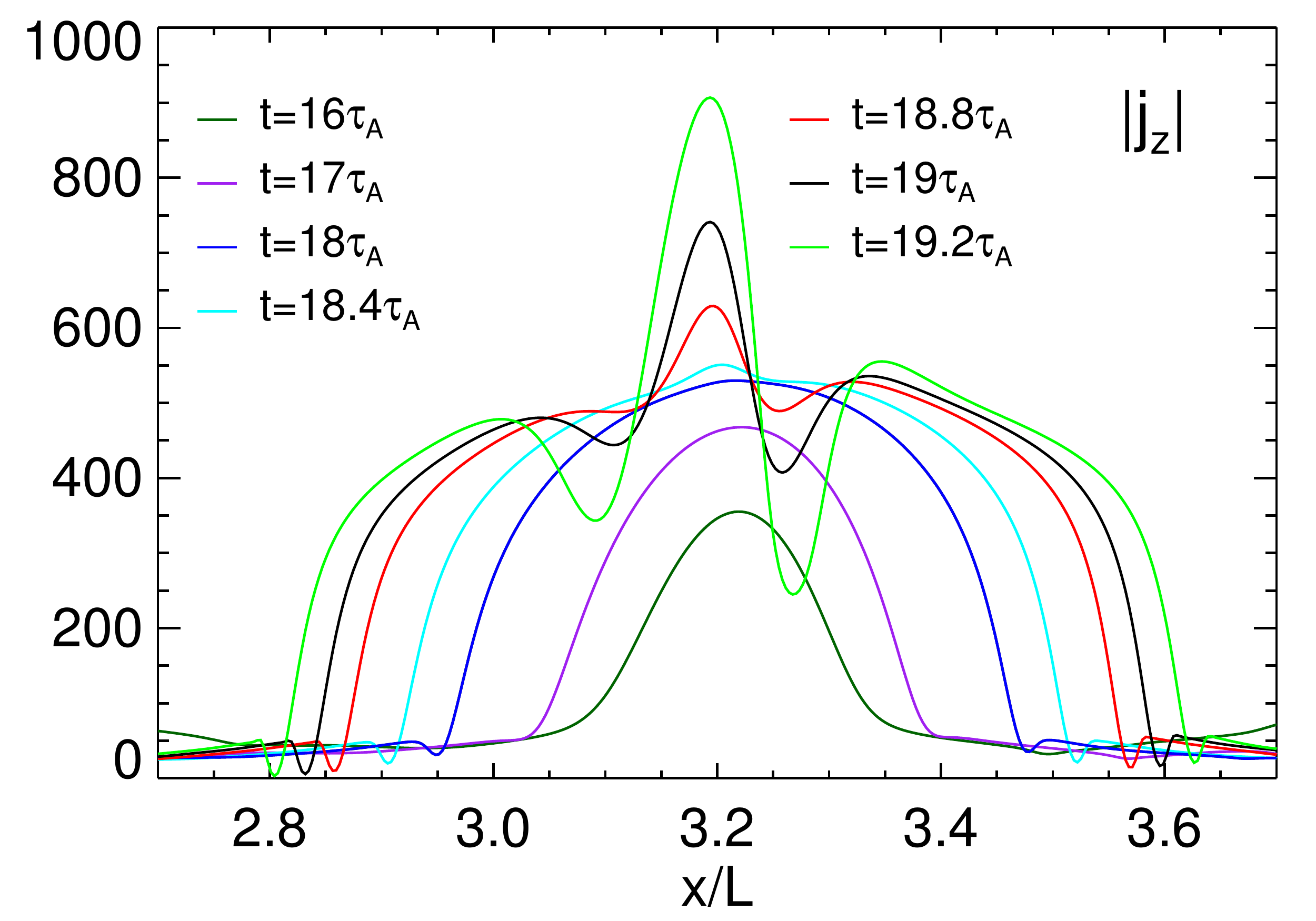}
\includegraphics[width=0.49\textwidth]{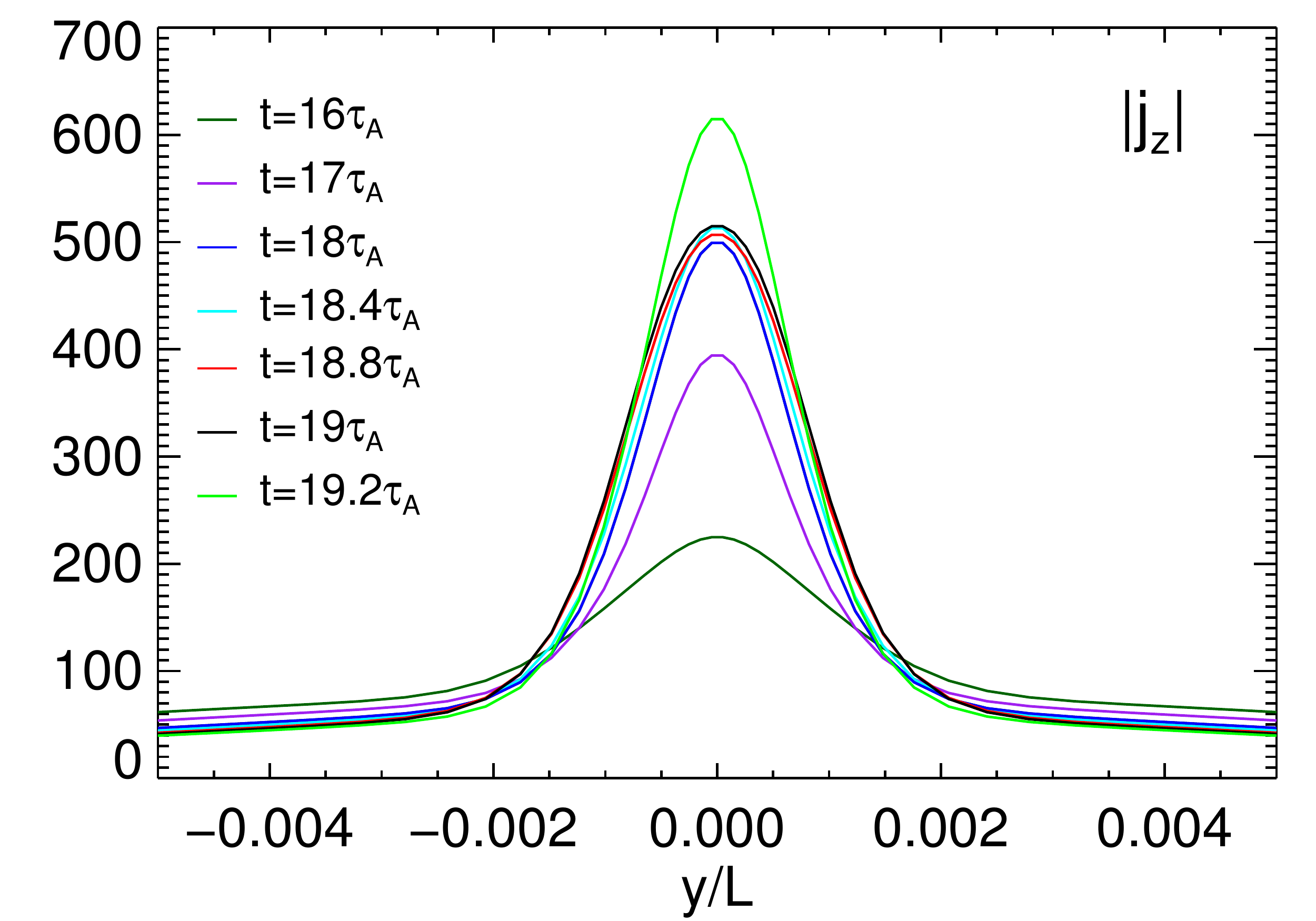}
\includegraphics[width=0.85\textwidth]{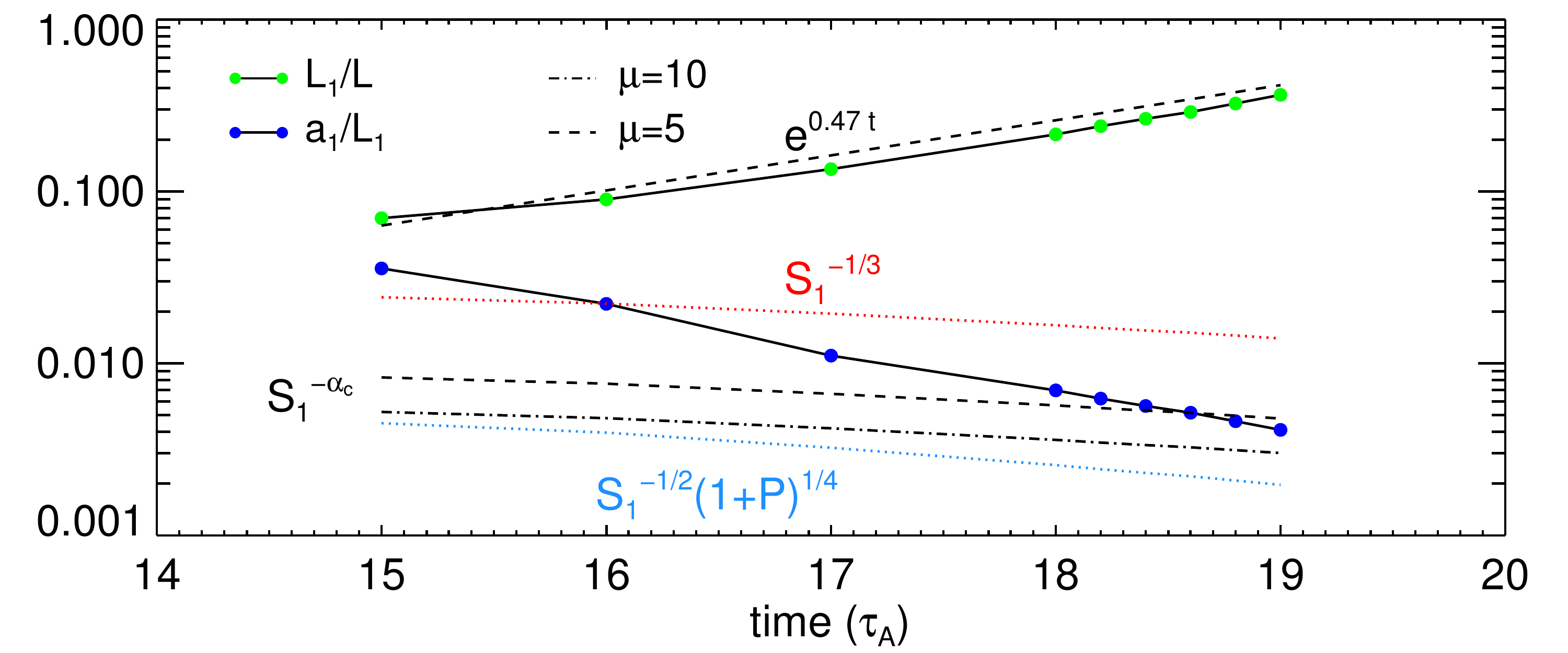}
\caption{{ Details of the time evolution of the secondary current sheet shown in Fig.~\ref{bolle}. Upper row: plot of the intensity} of the out-of-plane current density $j_z$ along (left panel, at $y/L=0$) and across ({ right panel}, at $x/L=3.15$) the sheet; the current half-length $L_1$ increases (exponentially) in time whereas its half-thickness $a_1$ is approximately constant, $a_1\simeq L\,0.0015$. {Lower panel (adapted from \citet{anna2}):  time evolution of the inverse aspect ratio $a_1/L_1$ (blue dots) and of the length $L_1/L$ (green dots) of the sheet; we plot for reference the thresholds given by the ``ideal" tearing (red dotted line), the viscous Sweet-Parker (light blue dotted line), and the corrected-flow ``ideal" threshold (dashed and dot-dashed black lines), the latter according to eq.~(\ref{outf})}.}
\label{jz}
\end{center}
\end{figure}

\begin{figure}
\begin{center}
\includegraphics[width=0.7\textwidth]{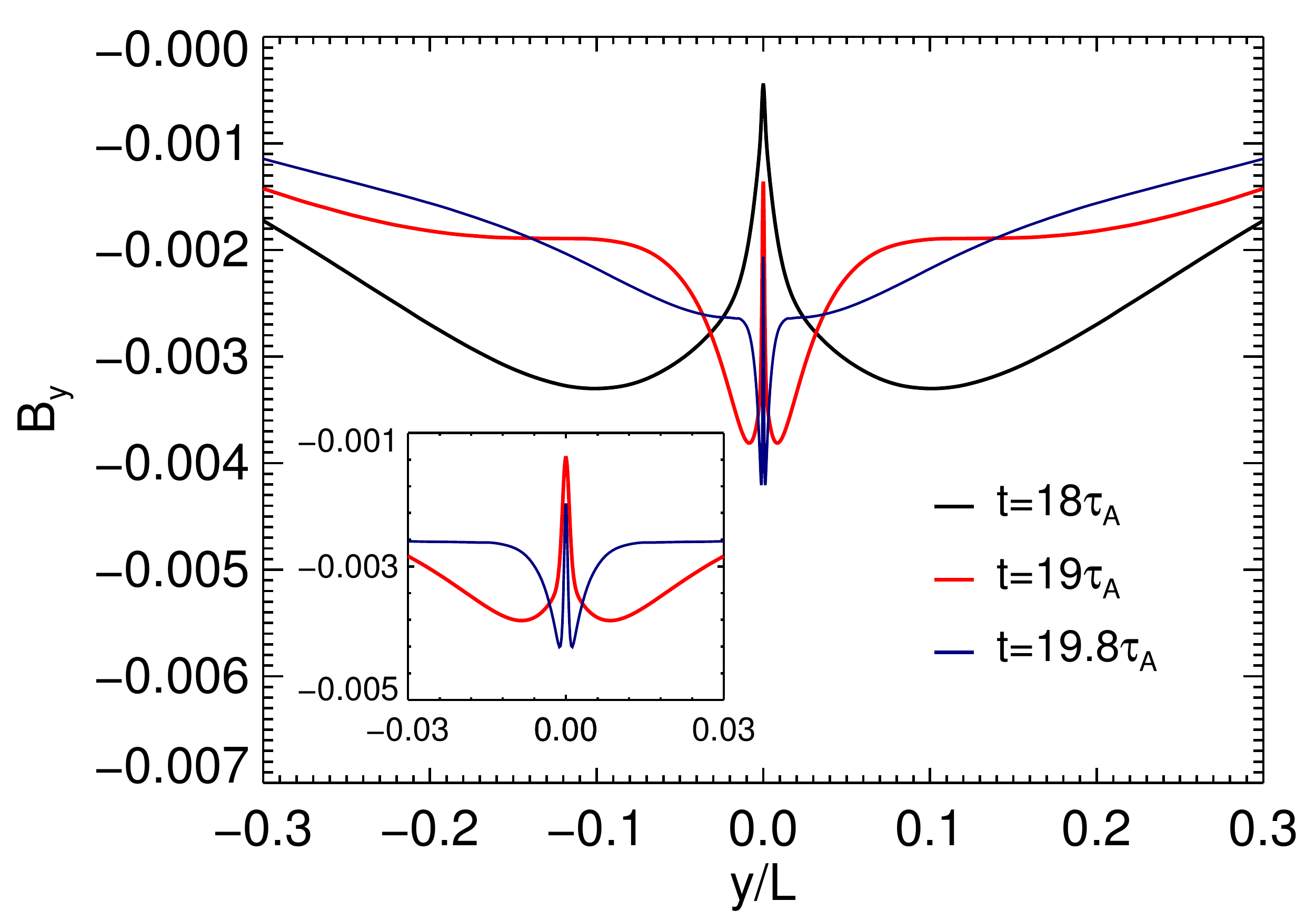}
\caption{Hierarchy of tearing modes: plot of the magnetic field component along the inhomogeneous direction, $B_y$, vs $y/L$ at position $x/L=3.15$ of the current sheet shown in Fig.~\ref{bolle}. Black color corresponds to the primary tearing, the red one to $n=1$, and the blue to $n=2$. The inset is a blow-up of the magnetic field for $n=1$ and $n=2$ (from~\cite{anna2}).}
\label{fractal}
\end{center}
\end{figure}

Recent numerical two-dimensional simulations of  ``ideally" unstable sheets at  Lund\-quist numbers $S=10^6-10^7$, have shown that the secondary current sheet lengthens exponentially at a rate close to the growth rate of the primary tearing~\citep{anna2}. { This is seen in  Fig.~\ref{jz}, where different panels describing the time evolution of the secondary current sheet shown in Fig.~\ref{bolle} are displayed. The (half) thickness and length of such a sheet is now labelled $a_1$ and  $L_1$: the upper left panel shows the profile of the  out-of-plane current density $j_z(x,y)$ intensity along the current sheet at the neutral line $y/L=0$ at different times, ranging from the end of the linear stage of the primary tearing up to the fully developed  secondary instability. The latter can be recognized as the growth of a more intense current maximum in the middle of $L_1$, surrounded by two local minima, corresponding to two plasmoids, at about $t=19.2\taua$ (green color); in the upper right panel we plot a cut of $j_z(x,y)$ across the sheet  at $x/L=3.15$, at the same times. Such a secondary current sheet arises from  the inner diffusion layer of the primary tearing: its  thickness has indeed an almost constant value $a_1/L\simeq0.0015$, that consistently scales with the macroscopic quantities as $a_1/L\simeq\delta_i/L\sim S^{-1/2}$ (see eq.~(\ref{id})). This can be seen from Fig.~\ref{jz}, upper right panel, for all times before the appearance of fully developed plasmoids (i.e. $t=19.2\taua$ as mentioned above); in the lower panel we plot the time evolution of $L_1$ (green dots), $a_1/L_1$ (blue dots), and the threshold conditions for instability corresponding to the ``ideal" tearing (red dotted line), to the viscous SP (light blue dotted line), and to the flow-modified ``ideal" tearing  discussed   in Section~\ref{sub_outf} (black dashed and dot-dashed  lines). As can be seen, this  current sheet  breaks-up during the collapse once it becomes unstable on the local Alfv\'en time, hence \emph{before} the SP aspect ratio is reached and in good agreement with the flow-modified ``ideal" tearing~\citep{landi_2015,anna2},  finally giving rise to another thinner sheet (see the green curves in Fig.~\ref{jz}, upper panels).}
In  Fig.~\ref{fractal} we show the profile of the magnetic field $B_y(y,x)$, $x/L=3.15$, across the center of the same  sheet shown in Fig.~\ref{bolle}, at three different times.  { The magnetic field profile displays a striking if unsurprising similarity to the tearing eigenfunctions (compare also with Fig.~\ref{modes}, left panel), as well as a hierarchical structure: the black color corresponds to the magnetic perturbation grown during the primary instability; the red and blue colors correspond to the first and to the second secondary instability, magnified  in the inset. In particular, note that the red profile as seen in the inset is the same as the black profile of the main figure. What this means is that each magnetic field perturbation that will lead to a new sheet grows within the inner diffusion layer of the unstable mode developed within the previous sheet. This in turn suggests that formation of ever smaller scales occurs via a recursive process of X-point formation, collapse and break-up, reminiscent of the so-called  ``fractal" reconnection scenario originally proposed by~\cite{shibata},  that can be accounted for by the ``ideal" tearing instability. 

Recursive models have been introduced for describing the different stages of such sub-layer formation (sometimes also referred to as the multiple X-point reconnection stage) and how many plasmoids are generated during the fully nonlinear evolution~\citep{dau_2009,huang_2010,uzdensky_2010,ji_2011}. To this end, it is convenient now to label with an $n$ the (half) length $L_n$, thickness $a_n$, and the local Lundquist number $S_n=(L_n/L)\,S$ of the $n$th current sheet. Previous models  assume that at each step current sheets become unstable giving rise to smaller ones with length $L_n=L_{n-1}/N_{n-1}$, where $N_{n}$ is the number of plasmoids at step $n$, and thickness defined by the SP scaling, therefore $a_n=L_nS_n^{-1/2}$. In particular, \cite{ji_2011} used a phenomenological scaling for the number of plasmoids $N_n=(S_n/S_c)^{\beta}$, where $\beta$ is left as a free parameter, and $S_c\simeq 10^4$ is the empirical critical Lundquist number for onset of  plasmoid instability; such a scaling describing  the number of plasmoids has been commonly adopted by other authors as well, motivated by the linear theory -- although, strictly speaking, the linear theory as developed does not predict a renormalization to $S_c$, and therefore it yields a far larger value for $N_n$. Those assumptions lead to an infinite hierarchy, where secondary current sheets never cross the presumed stability threshold $S_n\lesssim S_c$, and for this reason \cite{ji_2011} impose a minimum number of plasmoids $N_{min}$ as a cut-off to find the maximum index of the hierarchy, $n_*$. Assuming also that {\it each} current sheet becomes unstable, they find that the total, final number $\mathcal N$ of plasmoids is $\mathcal N\simeq(S/S_c)^{z}$, with $z\sim0.76,\,0.96$ for $n_*\simeq3,\,2$ and $\beta=3/8,\,0.8$. This heuristic argument then leads to values for the number of plasmoids consistent with that  found directly in numerical simulations~\citep{dau_2009,cassak_2009,huang_2010}.} 

Before illustrating our recursive model, it is worth commenting on the question of the scaling of the number of islands with the Lundquist number as the plasmoid instability develops. Numerical simulations appear to have shown a number compatible with the scaling of the plasmoid instability on Sweet-Parker sheets, though generally speaking simulations tend to have Lundquist numbers which, already at the macroscopic scale, are quite close to the critical one for stability threshold ($S_c$)~\citep{batta_2009,huang_2010}. In this sense it is surprising that a scaling relationship such as $N_{\text{\sc{sp}}}\sim S^{3/8}$, derived under the assumptions that flows may be completely neglected, might hold true. The same might be said of the scaling of islands along the sheet once nonlinearities become important, where results closer to $\mathcal N\sim S$ are found~\citep{huang_2010}. The reason is that counting X-points is equivalent to precisely determining the topology of the magnetic field at very small scales, while the (maximum possible) value of the Lundquist number $S$ is essentially determined by the number of grid points. A scaling  $\mathcal N\sim S$ along an extended one-dimensional sheet implies that the number of islands grows linearly with resolution. In such a reconnection configuration, the line is the original central current sheet neutral line: when moving to more general two-dimensional configurations, very high resolution simulations of fully developed turbulence show a number of X-points appearing to scale as $\mathcal N\sim S^{3/2}$~\citep{wan_2013}.  Following our previous statement, the corresponding  upper limit set by resolution on the scaling of the number of island  would be $\mathcal N\sim S^2$, so it appears that the scaling found by~\citet{wan_2013} is reasonable. In other words, as stated by \citet{wan_2013} ``[\dots] lack of adequate numerical resolution can easily increase the number of detected X-points, thus producing nonphysical results. Generally speaking, one requires high spatial resolution, to at least three times the Kolmogorov dissipation wavenumber, when using the pseudo-spectral approach that we have employed." 

Here we reconsider the recursive reconnection model, but following more in detail the ``ideal" tearing scenario guided by our numerical results. We point out that a hierarchy of sub-layers form anywhere  in regions where currents have room to collapse. In this regard, boundary conditions may play a fundamental role, if the  system size is not large enough to allow the dynamics to develop freely. Nevertheless, this limitation is automatically overcome in simulations of ideally unstable current sheets, since the wavelength of the ideal mode, $\lambda_i\sim 2\pi\,S^{-1/6}$, is much smaller than the length of the current sheet (approximately given by the size of the simulation domain). Of course, the collapse of X-points  can not and does not occur uniformly along the sheet, rather, nonlinear evolution is determined by the competition and interplay between  multiple recursive X-point formation within the most intense currents, and an inverse cascade of merging plasmoids, as can also be seen by inspection of Fig.~\ref{coll}. Therefore, differently from the models discussed above, we apply the recursive idea to a \emph{given} secondary current sheet, so that our model leaves the final number of plasmoids generated undefined.  It instead provides a prediction of the total number of steps $n_*$ and of the timescale of the recursive process itself. 

\begin{figure}
\begin{center}
\includegraphics[width=0.49\textwidth]{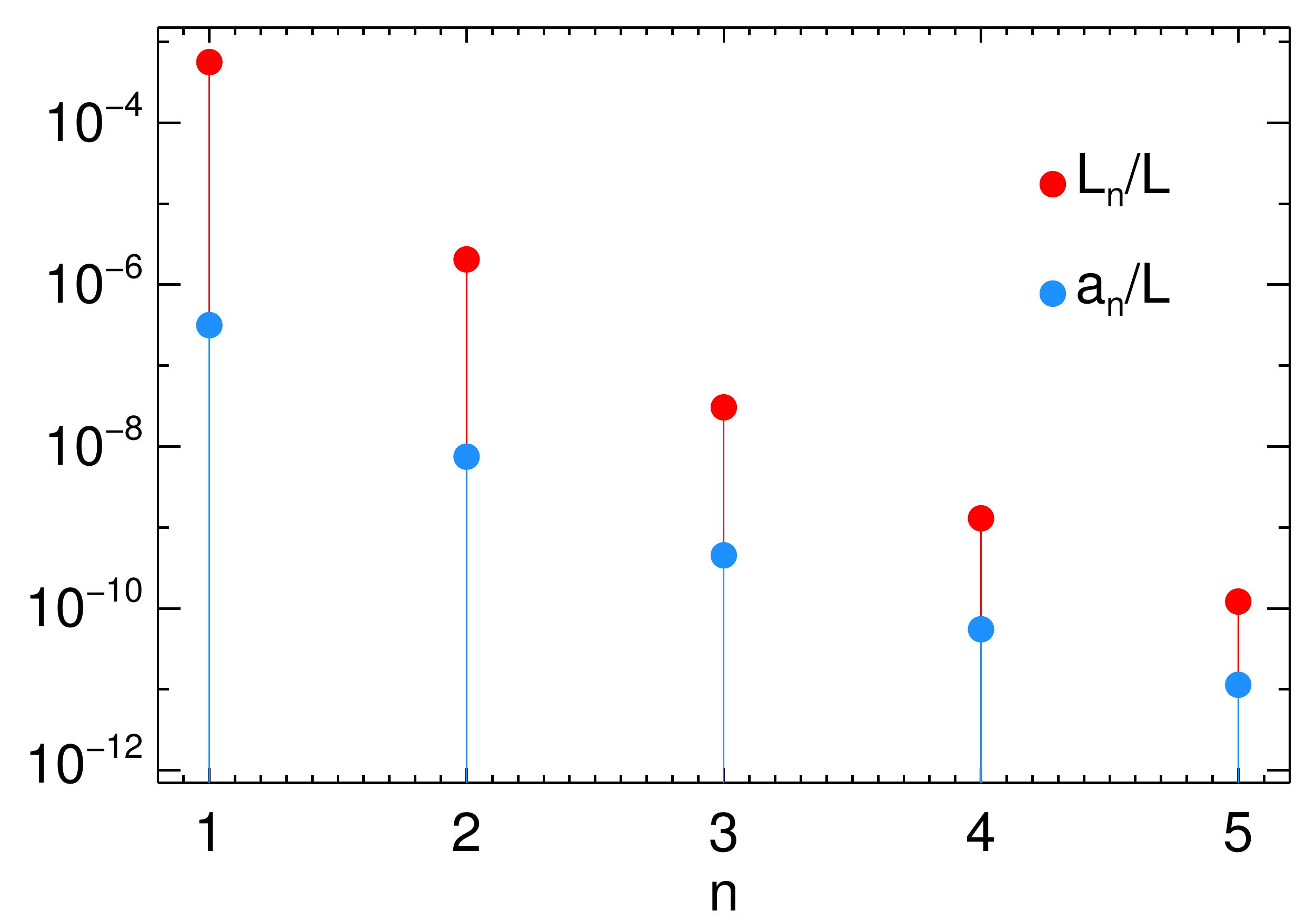}
\includegraphics[width=0.49\textwidth]{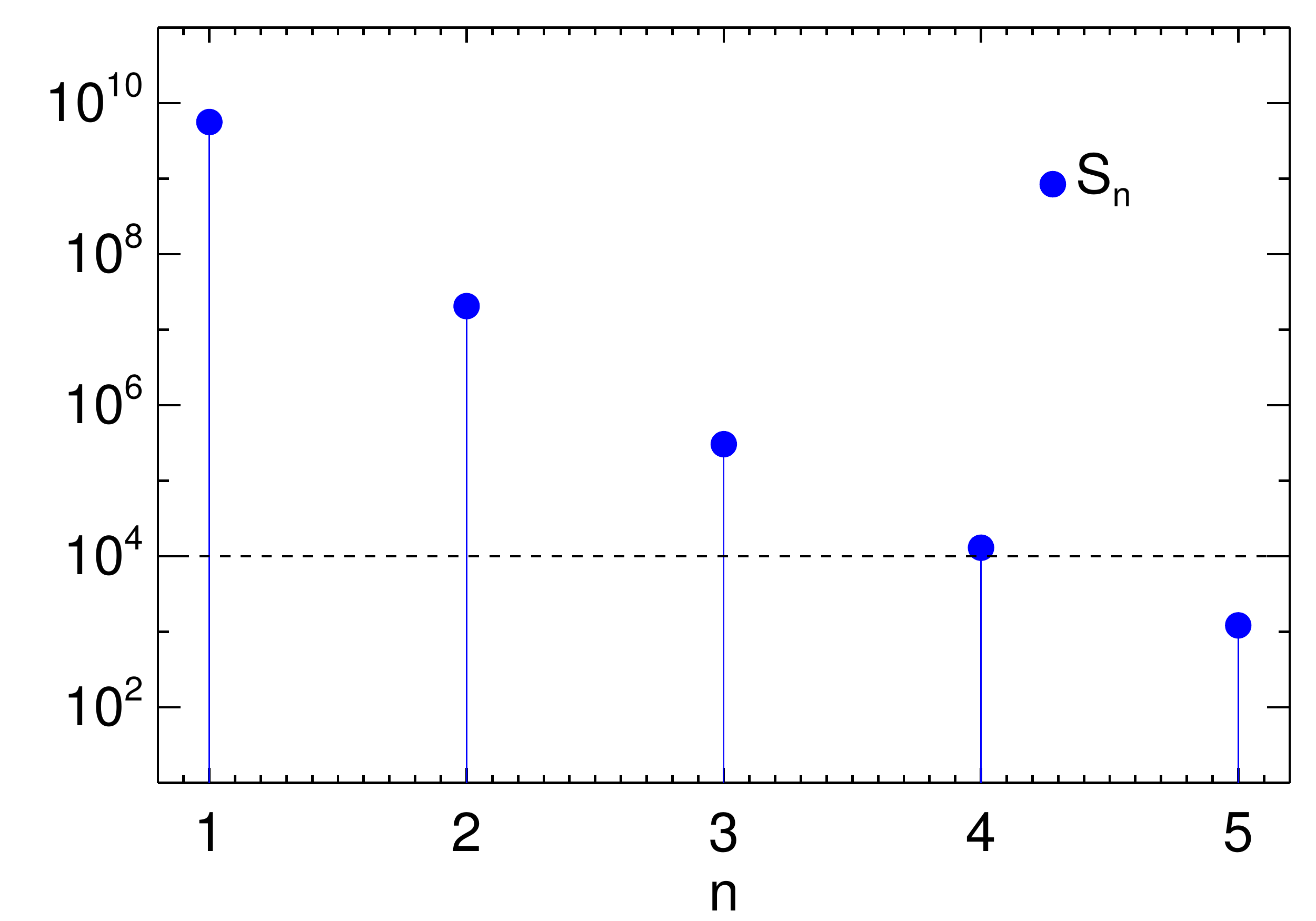}
\caption{Plot of the sequence $L_n/L$ (left) and $S_n$ (right) for $S=10^{13}$, see  eq.~(\ref{power}).}
\label{cascade}
\end{center}
\end{figure}

Since tearing is a multi-scale process, in the sense that it involves the formation of an inner singular layer, we also introduce the thickness of such  inner layer $\delta_n$ of the $n$th  unstable current sheet (for us $n\geq1$), and we define the local Alfv\'en time $\tau_{\text{\sc a},n}=(L_n/L)\,\taua$. {We rely on the following assumptions based on observations of our simulation results: first, that current sheets are lengthening and that they become unstable when the local critical aspect ratio is reached; second, we do not make any assumption on the length $L_n$, as is done  in other recursive models~\citep{shibata,ji_2011}, but, on the contrary, we observe that the thickness $a_n$  corresponds to the inner diffusion layer of the $(n-1)$th  tearing, $\delta_{n-1}$.} These two requirements translate respectively~into 
\begin{equation}
\frac{a_n}{L_n}\sim S_n^{-1/3},\,\,\frac{a_n}{L_{n-1}}\sim S_{n-1}^{-1/2},
\label{cond}
\end{equation}
where for simplicity we have neglected the stabilizing effect of  viscosity and of plasma flows, but which should be retained for the lower values of $S_n$ {(see Section~\ref{sub_outf})}.  Expressions given in~(\ref{cond})  yield the following scaling laws:     
\begin{equation}
L_n=L\,S^{-1+(3/4)^n},\,\, \tau_{\text{\sc a},n}=\frac{L_n}{L}\,\taua,\,\, S_n=S^{(3/4)^n}.
\label{power}
\end{equation}
Considering typical coronal conditions, for which $L\simeq10^9$~cm, $B_0\simeq50$~G, the number density $n_0\simeq10^9$~cm$^{-3}$, and the temperature $T\simeq10^6$~K, then the macroscopic Lundquist number is $S\simeq10^{13}$~\citep{brag}. The sequences of $L_n$ and of $S_n$  for $S=10^{13}$ are represented in Fig.~\ref{cascade},  showing that  the formation of microscopic scales occurs very rapidly.  In this case, after a  number of steps $n_*\simeq4$ the local Lundquist number reaches $S_{n_*}\simeq10^4$, {the region close to where one expects to find complete flow-driven stabilization.} The time required to reach this marginally stable state gives an indication of the timescale for the complete disruption of the original current sheet, and is given by the time required to trigger the first instability $\tau_0$ ($\tau_0\simeq\tau_i$) plus the time of the recursive reconnection, that for $S=10^{13}$ is about
\begin{equation}
\tau\simeq\tau_0+\sum_{n=1}^{n_*}\tau_{\text{\sc a},n}\simeq\big(10-100)\taua+5\times10^{-4}\taua.
\end{equation}  

Formation of smaller scales can mediate the transition to a Hall (or kinetic) reconnection regime, when the ion-scales are reached dynamically~\citep{cassak_2005,dau_2009,shepherd_2010,huang_2011}. If ion-scales are formed during the recursive reconnection, e.g., $a_n\simeq d_i$,  then  different scalings should  be adopted from that moment on. For instance, with the parameters chosen above as representative of the solar corona,  the normalized ion inertial length turns out to be about $d_i/L\simeq10^{-7}$, which is of the same order of the thickness formed at the first step, $a_1/L$, for $S=10^{13}$. A study including the Hall effect is however necessary to assess with more precision at which thickness the MHD scalings used so far are modified. { The main point here is that regardless of whether MHD or a specific kinetic regime is the more suitable model, ``ideal" tearing provides a clear well-defined threshold for the onset of fast reconnection~\citep{ideal_de}, so that eqs.~(\ref{power}) can be extended to kinetic regimes as well.}

\section{Comparison with previous models}
In spite of the different initial conditions, the evolution of the current sheet formed during the nonlinear stage of the ``ideal" tearing, as shown for instance in Fig.~\ref{bolle}, is qualitatively reminiscent of, and can be compared to the ``embedded reconnection" scenario~\citep{shay_2004,cassak_2009}, in which the  diffusion layer is embedded within a thicker underlying  current sheet. The argument of the embedded layer led~\cite{cassak_2009} to take into account the increasing Alfv\'en speed due to pile-up just upstream the inner diffusion layer, and to find in this way the scaling $a_1/L_1\sim S_1^{-1/3}$ for onset of instability (in order to avoid confusion,  we adopt here our notation, and $S_1=\va L_1/\eta$).  Nevertheless,  their scaling has a different origin and it stems from the criterion to destabilize a SP sheet, when  the local Lundquist number exceeds the  value $S_c=10^4$. \cite{cassak_2009} assumed that the diffusion layer ultimately becomes  an unstable SP sheet,  which however satisfies $a_1/L_1\sim S_{up}^{-1/2}$, $S_{up}$ being  the upstream Lundquist number, defined through the upstream magnetic field $B_{up}$ and $L_1$; $B_{up}$ can be approximated near the neutral line by linearizing the magnetic field, hence $B_{up}\simeq B_0(a_1/a)$. Direct substitution of $B_{up}$ into the SP scaling for the inverse aspect ratio yields  the same exponent $\alpha=1/3$ for onset of instability. On the other hand, the ``ideal" tearing scaling derives from requiring an $S$-independent growth rate from the complete eigenmode analysis of tearing instability, and the presence of flows modifies the scaling exponent $\alpha$ according to eq.~(\ref{outf}). For their values $S_1\simeq2.5\times10^4-10^5$ the exponent $\alpha_c\simeq0.48-0.46$, that give an unstable current sheet when $a_1/L_1\sim S_1^{-\alpha_c}\simeq 0.007-0.005$. These estimates are in agreement with the unstable inverse aspect ratios found in the~\cite {cassak_2009} paper. 

It is interesting to note that the scaling exponent $\alpha=1/3$ as an instability criterion was also derived by~\cite{shibata} in the framework of  ``fractal" reconnection, introduced for the first time by those authors as a possible phenomenological model for  current sheet disruption during solar flares. Based on the argument of~\cite{bisk_1986},  \cite{shibata} searched for a condition to overcome the stabilizing effect of flows on the tearing instability. They  concluded that, given a current sheet of inverse aspect ratio $a_n/L_n$, the maximum growth rate of the tearing had to {\it exceed} the plasmoid evacuation rate $\Gamma_n\simeq\va/L_n=1/\tau_{\text{\sc a},n}$ for the tearing to develop (in the end, effects of flows on the growth rate are not taken into account in their model). \cite{shibata} can find in this way  that the current sheet must {\it at least} have an inverse aspect ratio $a_n/L_n\leq S_n^{-1/3}$ in order to become unstable. Again, the  result of \cite{shibata}  stems from a different line of thought, as they searched for a maximum growth rate larger than the inverse of the ideal time $\tau_{\text{\sc a},n}$, and not for an ``ideal" growth rate,  independent from the Lundquist number.   Next, in their ``fractal" model the authors  assume that the length for each sub-layer corresponds to the wavelength of the fastest growing mode, i.e.,  $L_n=2\pi/k_{n}$, that yields  a different scaling law for $L_n$ with respect to ours, having a weaker dependence on~$n$.

\section{Summary and discussion}

We have given an overview of linear  tearing instabilities within thin current sheets at arbitrary aspect ratios in resistive MHD and discussed some extensions, namely, Reduced MHD with inclusion of kinetic effects. Theory shows that fast reconnection, developing on timescales independent of the non ideal terms in Ohm's law, sets-in at a critical aspect ratio $L/a_i$ that satisfies specific scaling-laws with plasma parameters such as $S$ and $d_e/L$. Such scaling-laws can be modified by viscosity at large Prandtl numbers, or by kinetic effects when the inner diffusion layer is of the order of, or smaller than, $\rho_s$ and $\rho_i$~\citep{pucci,anna1,ideal_de}. The main result that emerges from these studies is that in the asymptotic limit $S\rightarrow\infty$ or $d_e/L\rightarrow0$ a violent transition from a stable state (reconnecting over infinitely long times) to an ideally unstable one has to occur when approaching the critical thickness $a_i/L$ from above. The onset of fast reconnection cannot take place within current sheets that are thicker than critical, for sufficiently large values of $S$: unstable modes at smaller aspect ratios, $L/a<L/a_i$, have growth rates whose values tend to zero in the asymptotic limit, whereas, precisely at $L/a_i$, the full dispersion relation $\gamma(k)$ becomes rapidly peaked around $k_m$ (cfr., for instance, Fig.~\ref{gam}, right panel), the ``ideal" mode being the only surviving one in the asymptotic limit;  ``ideal" tearing onset, by breaking-up the current on an ideal timescale, prevents the spontaneous formation of much larger aspect ratio current sheets, such as  the Sweet-Parker one. Fully nonlinear MHD simulations show that the inner diffusion layer of the ideally reconnecting current sheet evolves during the nonlinear stage into an elongated secondary current sheet, that in turn becomes unstable to ``ideal" tearing at the local Alfv\'en time. This process of current sheet formation from of the inner diffusion layer proceeds in a recursive way at smaller and smaller scales, that can be accounted for by  properly rescaling ``ideal" tearing to the local length of the sheet~\citep{landi_2015, anna2},  even though an appropriate statistical study of the fully nonlinear stage has not been completed yet because of the incredibly high resolutions required to resolve the recursive formation of plasmoids down to the dissipative scales. 

In the solar corona, where inter-species collisions usually provide the dominant dissipation mechanism for reconnection, the Lundquist numbers are $S\simeq10^{12}-10^{14}$. For instance, for $S=10^{13}$ the critical inverse aspect ratio is  $a_i/L\sim S^{-1/3}\simeq 5\times10^{-5}$, i.e., for a loop structure of length $L\simeq10^9$~cm the critical thickness would be $a_i\simeq 500$~m, with an inner layer of about $\delta_i\simeq3$~m, intermediate between the ion inertial length, $d_i\simeq10$~m, and the ion Larmor radius, $\rho_i\simeq10$~cm. Two-fluid effects related to the Hall term in the generalized Ohm's law need therefore to be investigated to give a more realistic description of ideally unstable current sheets. Inclusion of such effects becomes however necessary when nonlinearities naturally lead to the formation of microscopic scales, e.g., during recursive reconnection. 

A different regime describes the Earth's magnetosphere, and in particular the magnetotail, for  which a kinetic description is more suitable~{\citep{coppi66,Schindler,vas,dau_1999,sitnov2014}}. Here,  resistive reconnection, due either to inter-specie collisions or to wave-particle scattering~\citep{coroniti}, is usually dominated by  reconnection induced by electron inertia. Assuming values of plasma density $n_0\simeq0.1$~cm$^{-3}$, electron temperature $T_e\simeq10^7$~K (for simplicity we consider ions and electrons at the same temperature) and magnetic field $B_0\simeq10^{-4}$G, and taking as typical sheet length $L\simeq10^9$~cm, then the Lundquist number ranges from $S\simeq10^{10}$ (for wave-particle scattering;~\cite{eastwood}) to $S\simeq10^{15}$ (for standard collisions), whereas $d_e^2/L^2\simeq 10^{-6}$. The electron skin depth is about $d_e\simeq 10$km and the ion length-scales are $\rho_i\simeq\rho_s\simeq300$~km, of the order of the ion skin depth, $d_i\simeq700$~km. Ionic scales are not far from the thickness of the central current sheet, that is observed to thin down to about 1000~km during the substorm growth phase, and for these parameters the kinetic eq.~(\ref{kin2}) predicts $a_i\simeq400$~km. However, what is the effective driving mechanism for collisionless reconnection, and which are the dominant kinetic effects to be retained, are still matter of debate and the models that we have discussed can provide only indicative estimates of the critical thickness there (recall also that they require a strong guide field, which is rarely observed in magnetotail). The collisionless model that we have considered includes some kinetic effects, namely, FLR corrections to a dominant gyrotropic dynamics, but it does not include Landau resonances and the electron pressure anisotropy, which instead are known to allow reconnection~\citep{coppi66,Schindler,vas,Cai}. Other kinetic effects relevant for the magnetotail configuration should be considered as well, including electron meandering orbits around the neutral line in the absence of strong a guide field~\citep{sonnerup_1971}, or particle temperature anisotropies which are known to affect reconnection rates~\citep{chen_palma,karimabadi,matteini_2013}.  In this regard, it is still unclear in which way more realistic equilibria having a magnetic field component parallel to the shear, that is, two-dimensional equilibria rather than a simple Harris sheet, impact the stability of the current sheet itself, and in particular the possible onset of kinetic tearing modes~\citep{sitnov2014,pritchett}. These problematics need to be further explored. 

In the present discussion we have not considered three-dimensional effects, which allow  new instabilities and may affect both the linear~\citep{dau_1999,baarlud_2012} and especially the nonlinear evolution~\citep{dal_2002,landi_2008,dau_2011} of tearing modes, allowing a much richer dynamical evolution than in two dimensions. Magnetic islands extend in the third direction giving rise to flux ropes, that in turn are subject to secondary ideal instabilities. If the guide field is not sufficiently strong~\citep{dal_2005}, flux ropes start to braid among themselves, by overcoming in the typical two-dimensional evolution characterized by magnetic island coalescence and X-point collapse. By favoring in this way secondary reconnection within multiple layers along the ropes, the nonlinear dynamics efficiently drives  the system into a turbulent state. It is therefore of interest to study how our onset scenario, but especially the nonlinear recursive reconnection model, change in three-dimensions.

\acknowledgements This work was supported by the NASA program LWS, grant  NNX13AF81G. MV acknowledges support via the NASA Solar Probe Plus observatory scientist grant. The authors thank the editors and R. Lionello and F. Pegoraro for their useful comments and suggestions. Computational resources
supporting this work were provided by the NASA High-End Computing (HEC) Program through
the NASA Advanced Supercomputing (NAS) Division at Ames Research Center.

\end{document}